\documentclass[runningheads]{llncs}
\usepackage{graphicx}
\usepackage{amssymb}
\usepackage{multirow}
\usepackage{wasysym}
\usepackage{enumitem}
\usepackage{xcolor}
\begin{document}
%
\title{DDR-Net: Dividing and Downsampling Mixed Network for Diffeomorphic Image Registration}
\titlerunning{DDR-Net}
%

\author{Ankita Joshi\inst{1} \and
Yi Hong\inst{2}}
\authorrunning{Ankita Joshi and Yi Hong}
%
\institute{Department of Computer Science, University of Georgia \\ \email{ankita.joshi25@uga.edu}   \and Department of Computer Science and Engineering, Shanghai Jiao Tong University \\ \email{yi.hong@sjtu.edu.cn} }
\maketitle              
\begin{abstract}
Deep diffeomorphic registration faces significant challenges for high-dimensional images, especially in terms of memory limits. Existing approaches either downsample original images, or approximate underlying transformations, or reduce model size. The information loss during the approximation or insufficient model capacity is a hindrance to the registration accuracy for high-dimensional images, e.g., 3D medical volumes. In this paper, we propose a Dividing and Downsampling mixed Registration network (DDR-Net), a general architecture that preserves most of the image information at multiple scales. DDR-Net leverages the global context via downsampling the input and utilizes the local details from divided chunks of the input images. This design reduces the network input size and its memory cost; meanwhile, by fusing global and local information, DDR-Net obtains both coarse-level and fine-level alignments in the final deformation fields.
We evaluate DDR-Net on three public datasets, i.e., OASIS, IBSR18, and 3DIRCADB-01, and the experimental results demonstrate our approach outperforms existing approaches. Codes are available --here--.



\keywords{Diffeomorphic image registration  \and Dividing and downsampling \and Multi-scale registration.}
\end{abstract}
\section{Introduction}
Deformable image registration establishes pixel- or voxel-level dense correspondences for 2D or 3D image pairs, which form a deformation that transforms images into a common space for comparison and analysis. Such a deformation desires a good property of diffeomorphism, a smooth transformation with a smooth inverse, to ensure the preservation of topology when warping images. Classical image registration models, e.g., LDDMM~\cite{beg2005computing}, Stationary Velocity Fields (SVF)~\cite{arsigny2006log}, successfully estimate diffeomorphic deformations for building dense correspondences between image pairs. However, these models face challenges for practical applications, i.e., providing both fast and accurate solutions. Therefore, researchers have been working on improving the efficiency of diffeomorphic image registration methods~\cite{ashburner2007fast,zhang2015finite}. 

Recently, deep learning based approaches open an alternative to address the above challenges, which motivates our work in this paper. Existing deep registration models focus on tackling the efficiency challenge using supervised~\cite{yang2017quicksilver} or unsupervised techniques~\cite{dalca2018unsupervised,krebs2018unsupervised}. Supervised approaches~\cite{yang2017quicksilver} maintain the diffeomorphic property, which is inherited from the classical diffeomorphic model LDDMM, but it requires extra effort to obtain the ground-truth deformations. Meanwhile, its registration accuracy is limited by that of the obtained deformations. The unsupervised approaches~\cite{dalca2018unsupervised,krebs2018unsupervised} have shown promising diffeomorphic and efficient registration results by introducing an integration layer into the network design, based on the scaling and squaring method~\cite{higham2005scaling}. Due to the flexibility in selecting the network architecture and the loss function, the unsupervised framework has the potential to further improve the registration accuracy. However, because the integration step is computationally expensive and the network faces the memory challenge for high-dimensional images, the unsupervised approaches often work on downsampled images or deformations, which does not fully leverage available information and limits the registration accuracy. 

\begin{figure}[t]
    \centering
    \includegraphics[width=1.0\textwidth]{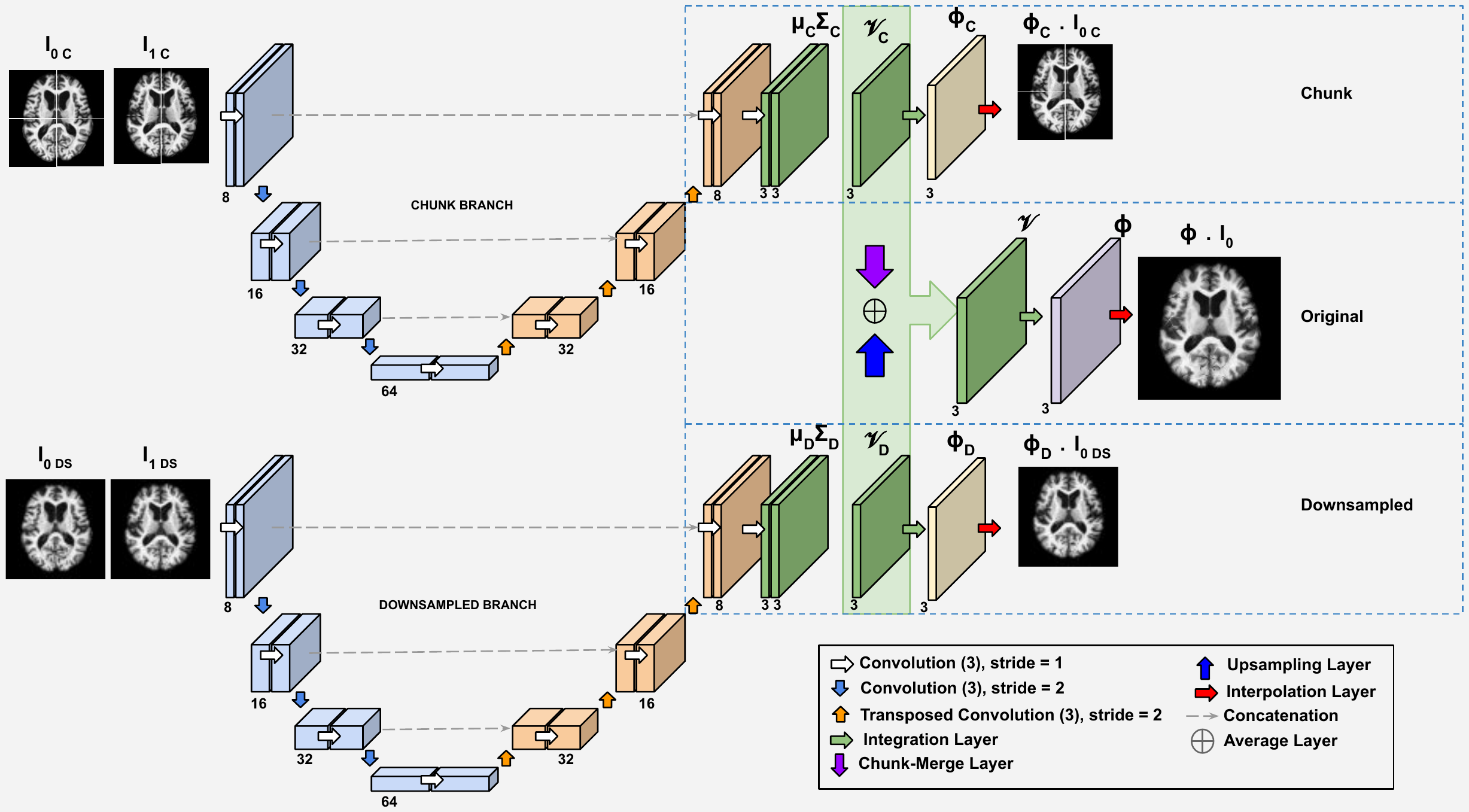}
    \caption{Architecture of our proposed DDR-Net. Given an image pair $I_0$ and $I_1$, the network estimates deformations at local scale, i.e., chunk branch, the global scale, i.e. the downsampled branch, and the original scale. The deformation $\phi$ at each level is driven by the corresponding velocity field $v$, which is sampled from the U-Net outputs, the mean $\mu$ and the variance $\Sigma$. The deformation at the original resolution is obtained by merging velocity fields generated by the global and local branches and then used to warp the original source images to the corresponding target images by using an interpolation layer. The number of the filters used are under the blocks.}
    \label{fig:overview}
\end{figure}

In this paper, we aim to improve the accuracy of unsupervised image registration, using a multi-scale design to integrate image deformations at global, local, and original scales. 
The difference in our work from current unsupervised registration models~\cite{dalca2018unsupervised,krebs2018unsupervised} is that our model works on downsampled images and chopped chunks, which reduce the computational and memory cost compared to working on the original image size directly. Meanwhile, the integration of these two-scale deformations back to the original scale improves the registration accuracy because of the information fusion at different levels. Previous work \cite{hering2019mlvirnet,krebs2019learning,yang2017quicksilver} either use multi-scale approaches of downsampling the entire image to improve accuracy of the result or use only patches of the images to reduce the memory cost, but both these techniques do not adequately leverage the data. Therefore, to gain an accuracy boost but not run into memory issues, we propose the Dividing and Downsampling mixed Registration Network (DDR-Net).

\vspace{0.05in}

\noindent
Our contributions in this paper are summarized as follows:
\begin{itemize}[noitemsep,topsep=0pt]
    \item We propose a novel architecture DDR-Net, which can effectively use both global and local features yielding high quality registration performance. Multi-scale information benefits the task of deep image registration.
    
    \item We demonstrate an effective way to obtain a trade-off between fully leveraging the available data under limited computing resources and gaining an improved accuracy of diffeomorphic image registration at the same time. 
    
    
    \item We conduct extensive experiments on both 2D and 3D datasets with different image types, including brain MRIs and liver CT scans. The experimental results demonstrate that our framework has better registration performance compared to deep-learning-based method VoxelMorph~\cite{dalca2018unsupervised} and the classical registration method ANTs SyN~\cite{avants2011reproducible}, in terms of image matching, deformation smoothness, and multi-structure segmentation.
\end{itemize}

\section{Dividing and Downsampling mixed Registration Network (DDR-Net)}

\noindent
\textbf{Architecture Overview.} As shown in Figure~\ref{fig:overview}, our proposed DDR-Net includes three main components: a global branch that handles the registration for downsampled images, a local branch that handles the registration for the cropped local chunks of original images, and an original branch that merges estimated velocity fields from the global and local branches to register original images. Each branch outputs a deformation field $\phi$ to register image pairs at its corresponding level, and they share some network designs as discussed below. 

\vspace{0.05in}
\noindent
\textbf{Backbone Registration.} At each scale, we have a diffeomorphic image registration problem. 
Given an image pair, a source image $I_{0}$ and a target image $I_{1}$, each of size $n_{x} \times n_{y} \times n_{z}$, the goal of diffeormorphic image registration is to estimate a smooth deformation field $ \phi: \mathbb{R}^{n_x \times n_y \times n_z} \rightarrow \mathbb{R}^{n_x \times n_y \times n_z} $  with a smooth $\phi^{-1}$, such that the image deformed from the source, i.e. $\phi \cdot I_{0}$, is similar to the target image $I_{1}$. Such a diffeomorphic deformation field is driven by a smooth velocity field $v_{t} , t \in [0,1]$, via the following differential equation:
\begin{equation}
    \frac{d}{dt}\phi = v_{t} \circ \phi_{t},  \quad  
    \phi_{0} = id.
    \label{eq:diff}
\end{equation}
Here, $id$ is an identity deformation. This formulation estimates an optimal velocity field $v$ that drives a deformation field $\phi$ to match an image pair. So, the registration network has three sub-tasks, i.e., estimating the velocity field, solving Eq.~(\ref{eq:diff}) for deformations, and deforming an image with interpolation. 

\vspace{0.05in}
\noindent
\textbf{Velocity Field Estimation.}
The global and local branches follow the same UNet~\cite{ronneberger2015u} architecture as shown in Fig.~\ref{fig:overview}. The UNet takes in image pairs and outputs the mean $\mu$ and the variance $\Sigma$ for sampling a corresponding stationary velocity field $v$. Here, the stationary velocity field assumption simplifies the solution of Eq.~(\ref{eq:diff}). Given a collection of image pairs $\{(I_0, I_1)\}$, where $I_0, I_1 \in \mathbb{R}^{n_x \times n_y \times n_z}$, the downsampling branch takes the low-resolution image pairs $\{(I_{0D}, I_{1D})\}$, downsampled by half, i.e., $I_{0D}, I_{1D} \in \mathbb{R}^{\frac{n_x}{2} \times \frac{n_y}{2} \times \frac{n_z}{2}}$. The chunk branch receives each input as $k$ divided patches with the same resolution as $I_{0D}$ and $I_{1D}$, i.e., $k \times\{(I_{0C}, I_{1C})\}$ and $I_{0C}, I_{1C} \in \mathbb{R}^{\frac{n_x}{2} \times \frac{n_y}{2} \times \frac{n_z}{2}}$. These cropped chunks have small overlaps on their boundaries to mitigate the discontinuity of the generated velocity fields at the chunk boundaries, when merging back to form the original high resolution velocity fields. The detailed network architecture in terms of the number of convolution layers and the kernel sizes are shown in Fig.~\ref{fig:overview}.


\vspace{0.05in}
\noindent
\textbf{Deformation Integration.} Under that assumption of the stationary velocity field, Equation~(\ref{eq:diff}) is simplified with a constant velocity field, and its solution is $\phi = e^{v}$. Similar to VoxelMorph, we adopt the scaling and squaring algorithm~\cite{higham2005scaling} to approximate this solution, which is implemented as a differentiable layer in the network. The downsampling and chunk branches integrate deformations separately, which are driven by their respective velocity fields. In the original branch, we use an averaged global and local velocity field to integrate the deformation. In particular, we upsample the global velocity field to the original resolution using an upsampling layer and merge the $k$ chunks to the original volume using a chunk-merge layer. As a result, we obtain a velocity field at the original resolution which is integrated to get the deformation at the original level. 

\vspace{0.05in}
\noindent
\textbf{Image Interpolation.} We use an interpolation layer to deform the source image at each branch. For each voxel $p$ in the target image, we compute its location $\phi(p)$ in the source image and compute its intensity value using linear interpolation. This differentiable operation allows the backpropagation of the network errors.

\vspace{0.05in}
\noindent
\textbf{Loss Functions.} 
All the velocity fields generated in the network have the constraint to ensure the smoothness of the deformations. Each branch also outputs a deformed source image to match the corresponding target image at each level, i.e., the downsampled, chunk, and original scales. We use the Mean Square Error (MSE) to measure the matching results and a K-L divergence loss to encourage the smoothness of the velocity field as done in \cite{dalca2018unsupervised,krebs2019learning}.


\begin{figure}[t]
    \centering
    \begin{tabular}{c|ccccl}
         Image Pair & $\phi \cdot I_0$ & $\phi \cdot I_0 - I_1$ &  $\phi$ &  $det(J_{\phi})$ \\
         \begin{tabular}{c}
            $I_0$ \\
            \raisebox{-.5\height}{\includegraphics[height=0.16\textwidth]{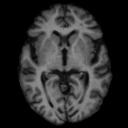}} \\ \\
            $I_1$ \\
            \raisebox{-.5\height}{\includegraphics[height=0.16\textwidth]{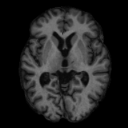}}
        \end{tabular}
       &
       \begin{tabular}{c}
             \raisebox{-.5\height}{\includegraphics[height=0.16\textwidth]{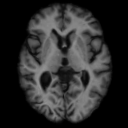}}\\ 
             \raisebox{-.5\height}{\includegraphics[height=0.16\textwidth]{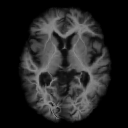}} \\
             \raisebox{-.5\height}{\includegraphics[height=0.16\textwidth]{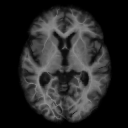}} 
       \end{tabular}
        &
        \begin{tabular}{c}
             \raisebox{-.5\height}{\includegraphics[height=0.16\textwidth]{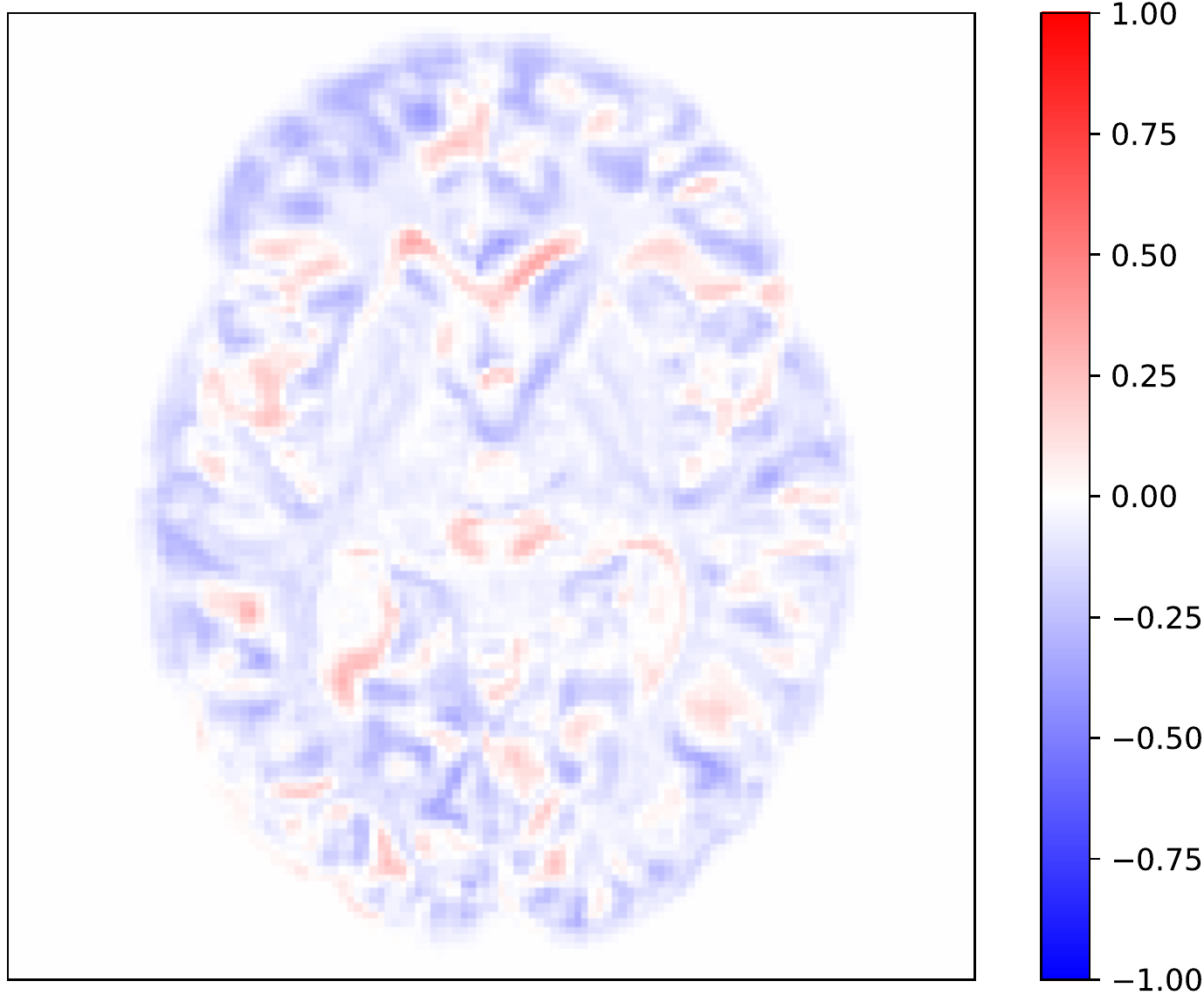}} \\ 
             \raisebox{-.5\height}{\includegraphics[height=0.16\textwidth]{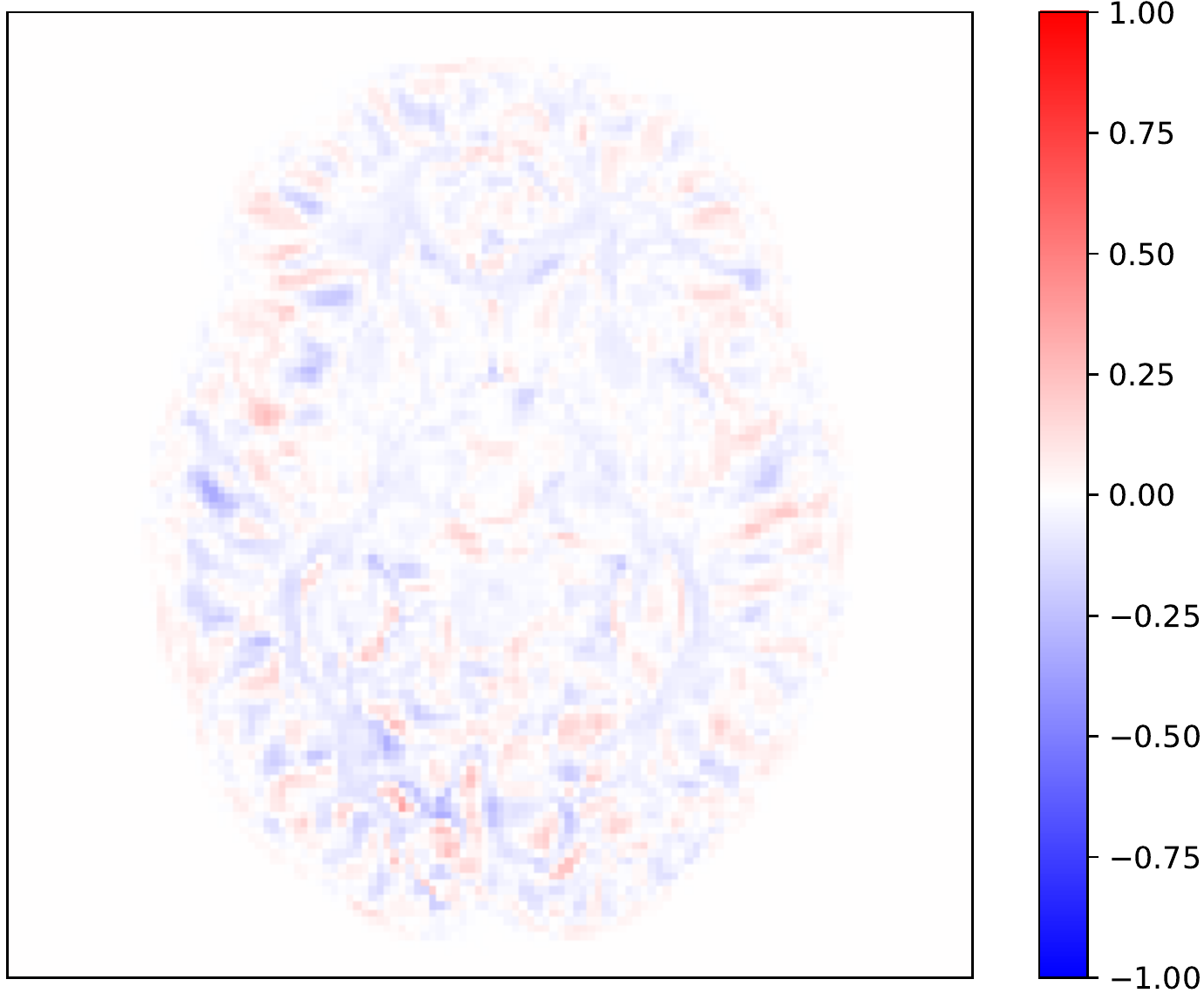}} \\
             \raisebox{-.5\height}{\includegraphics[height=0.16\textwidth]{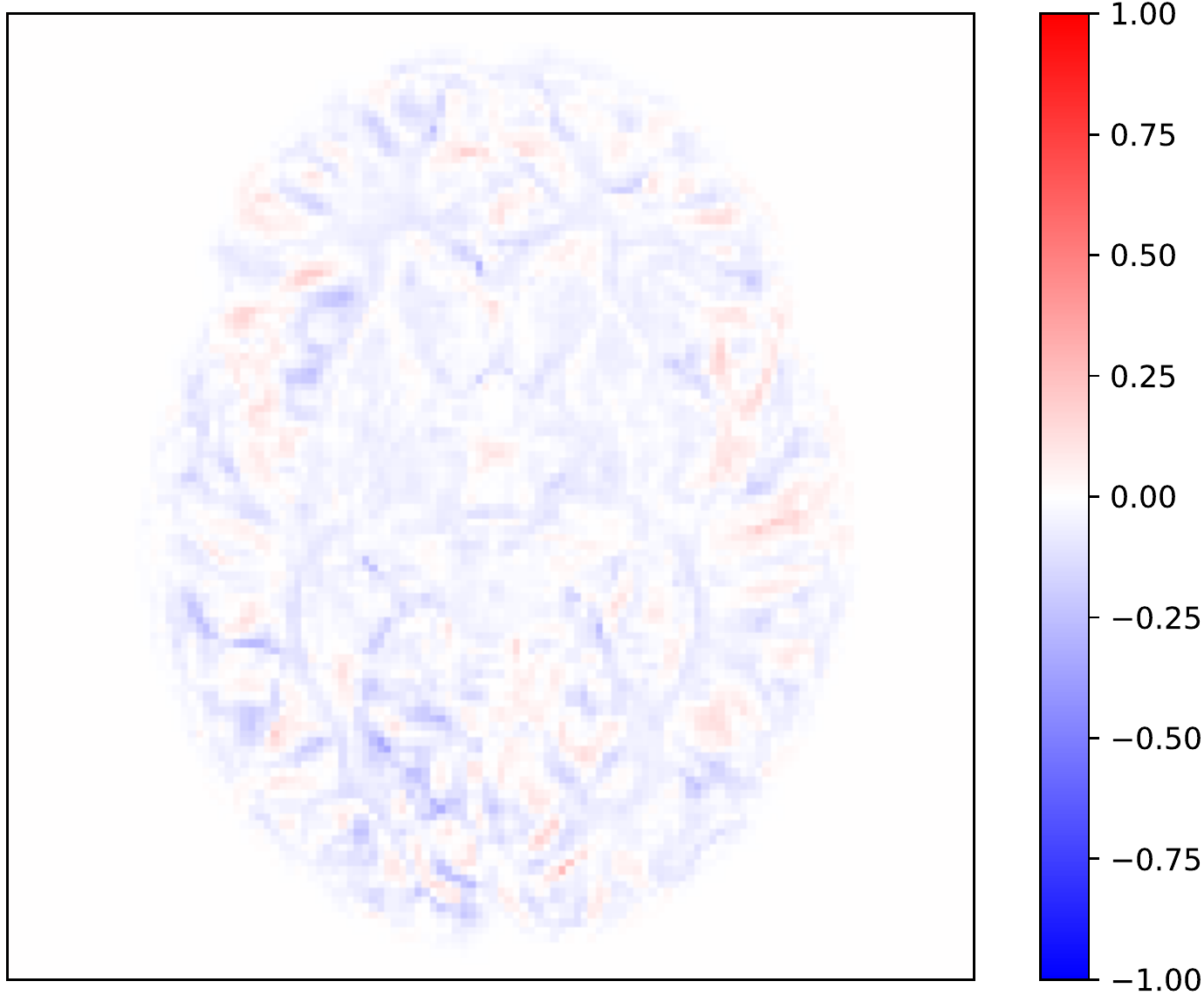}}  
         \end{tabular}
       &
        \begin{tabular}{c}
             \raisebox{-.5\height}{\includegraphics[height=0.16\textwidth]{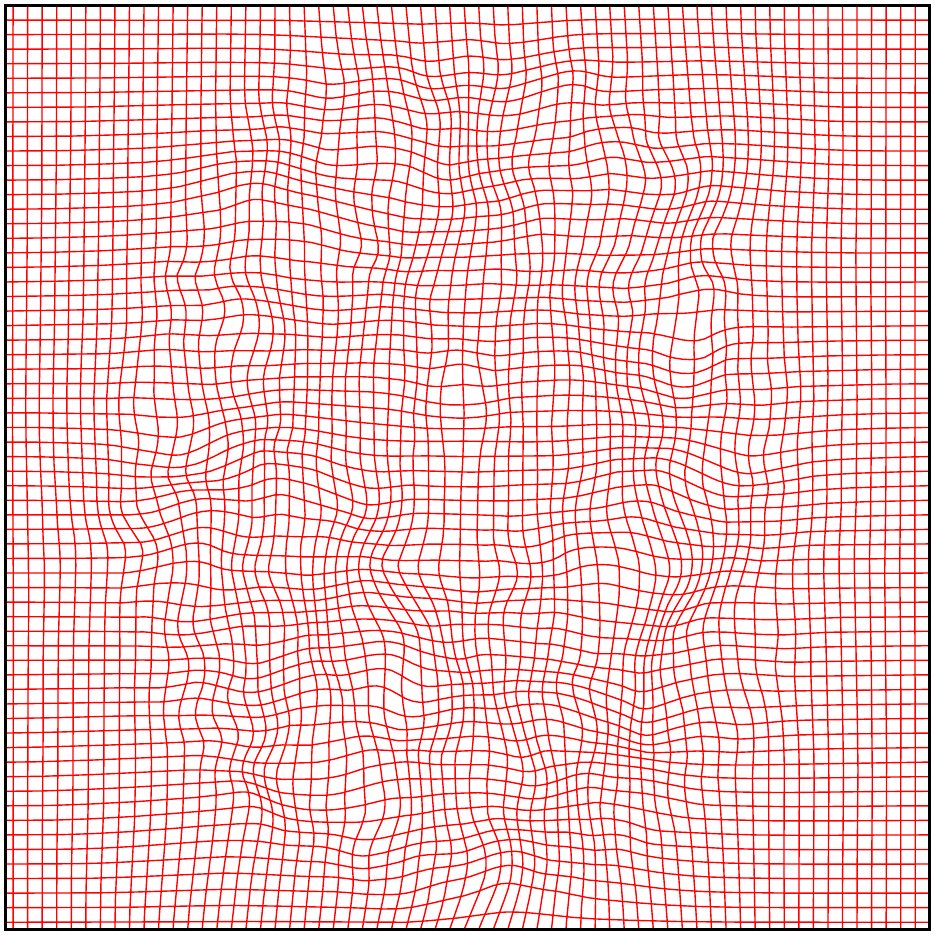}} \\ 
             \raisebox{-.5\height}{\includegraphics[height=0.16\textwidth]{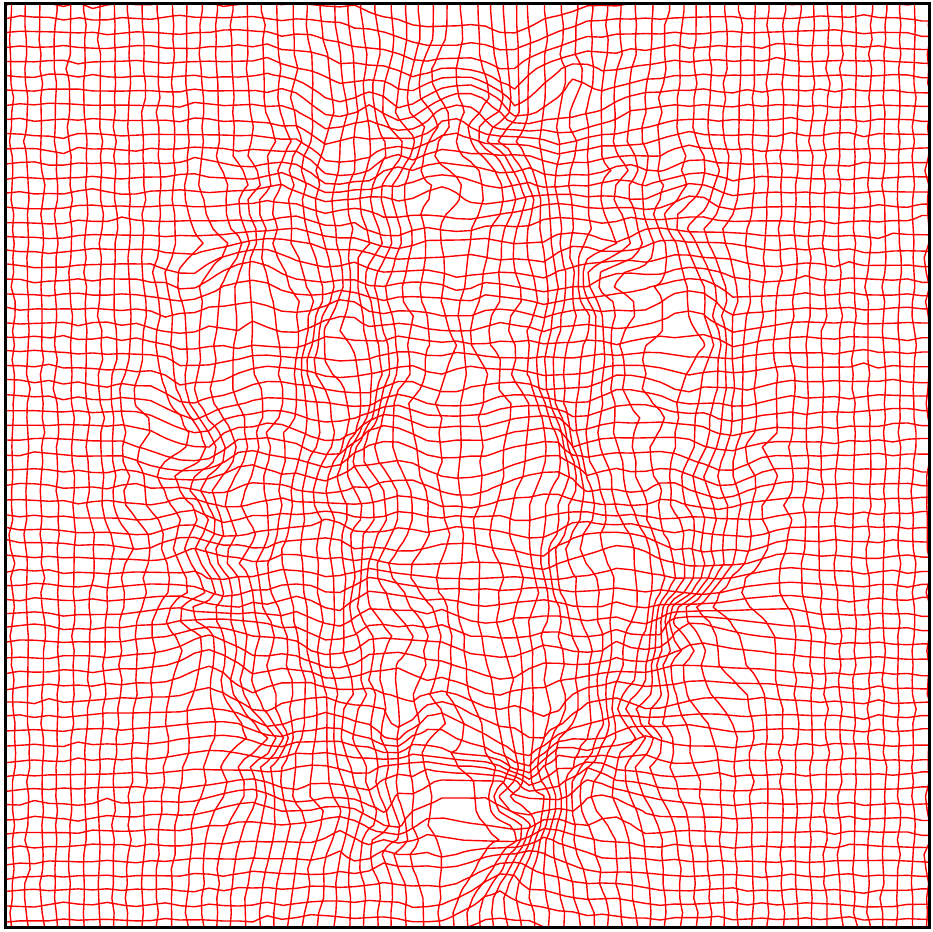}} \\
             \raisebox{-.5\height}{\includegraphics[height=0.16\textwidth]{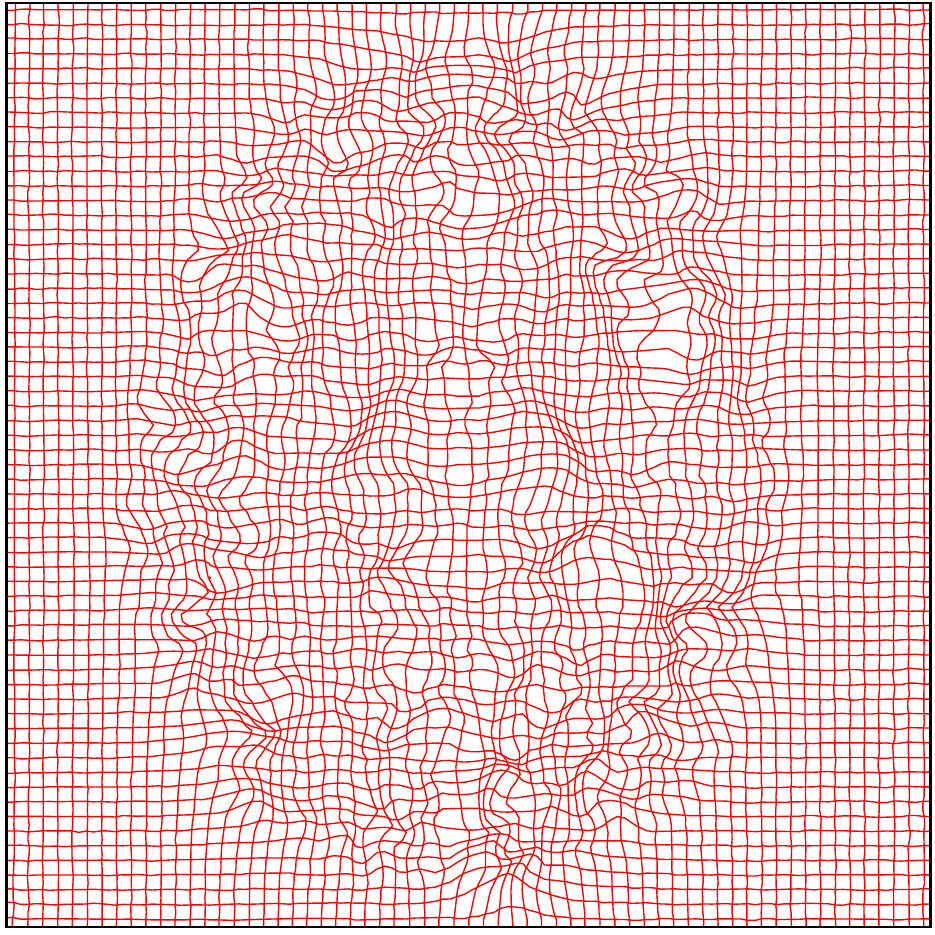}} 
         \end{tabular}
       
       &
       \begin{tabular}{c}
             \raisebox{-.5\height}{\includegraphics[height=0.16\textwidth]{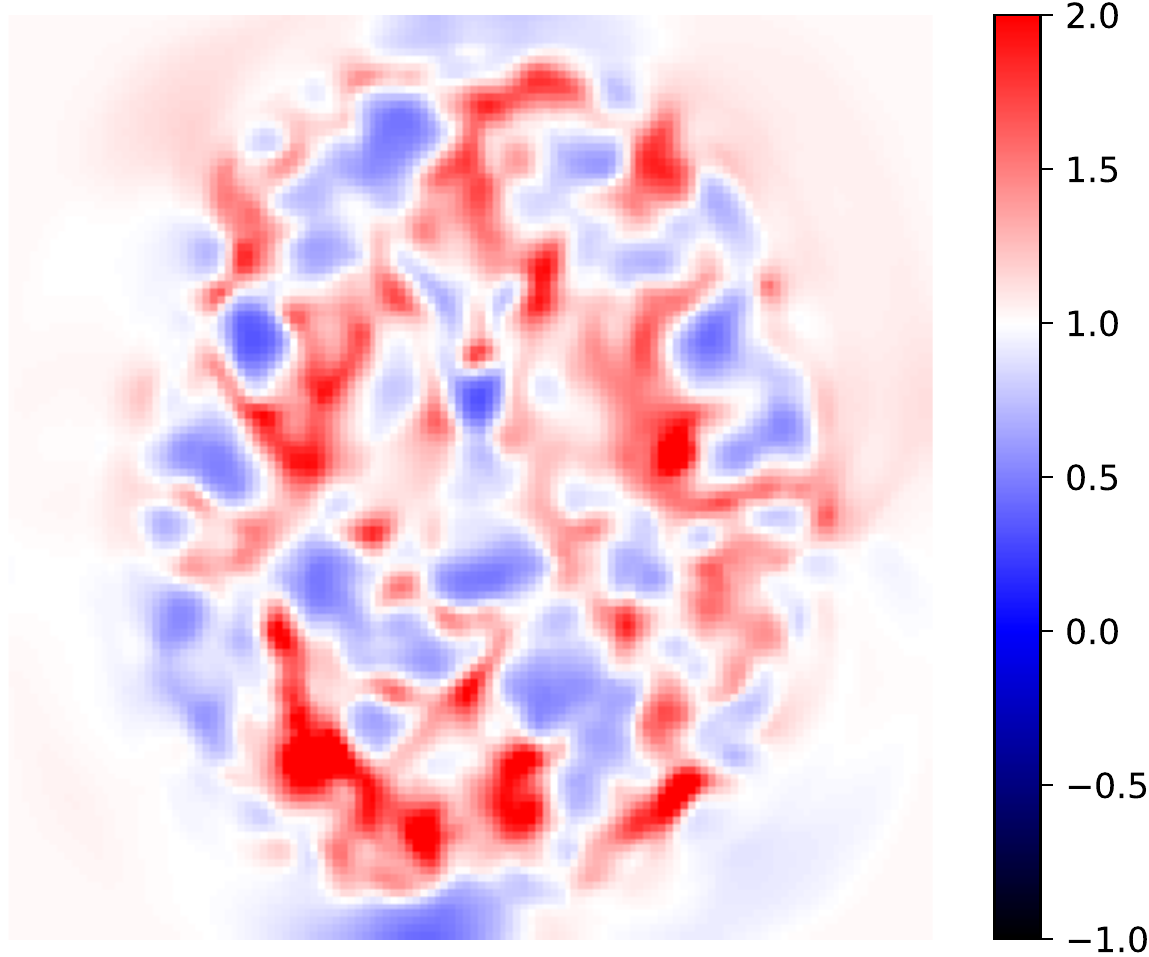}} \\ 
             \raisebox{-.5\height}{\includegraphics[height=0.16\textwidth]{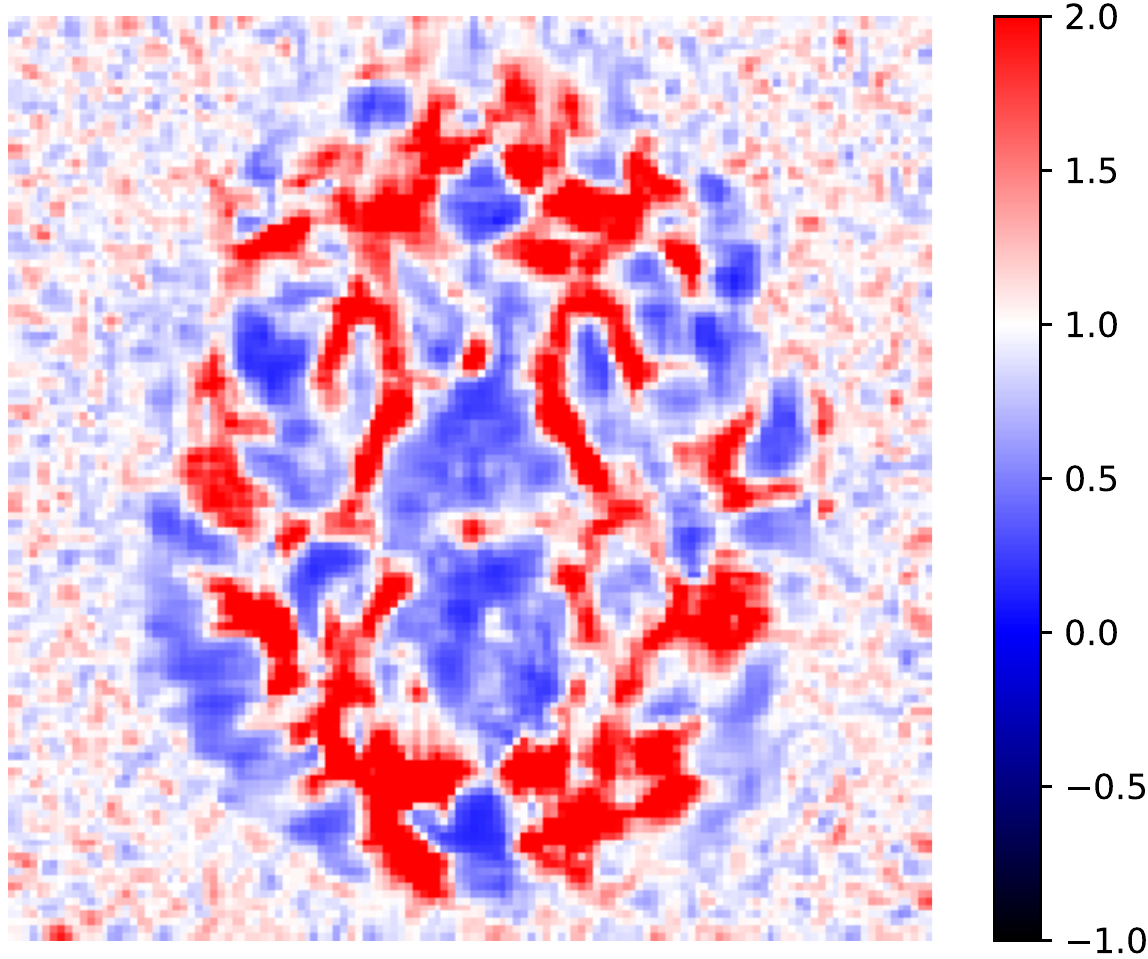}} \\
             \raisebox{-.5\height}{\includegraphics[height=0.16\textwidth]{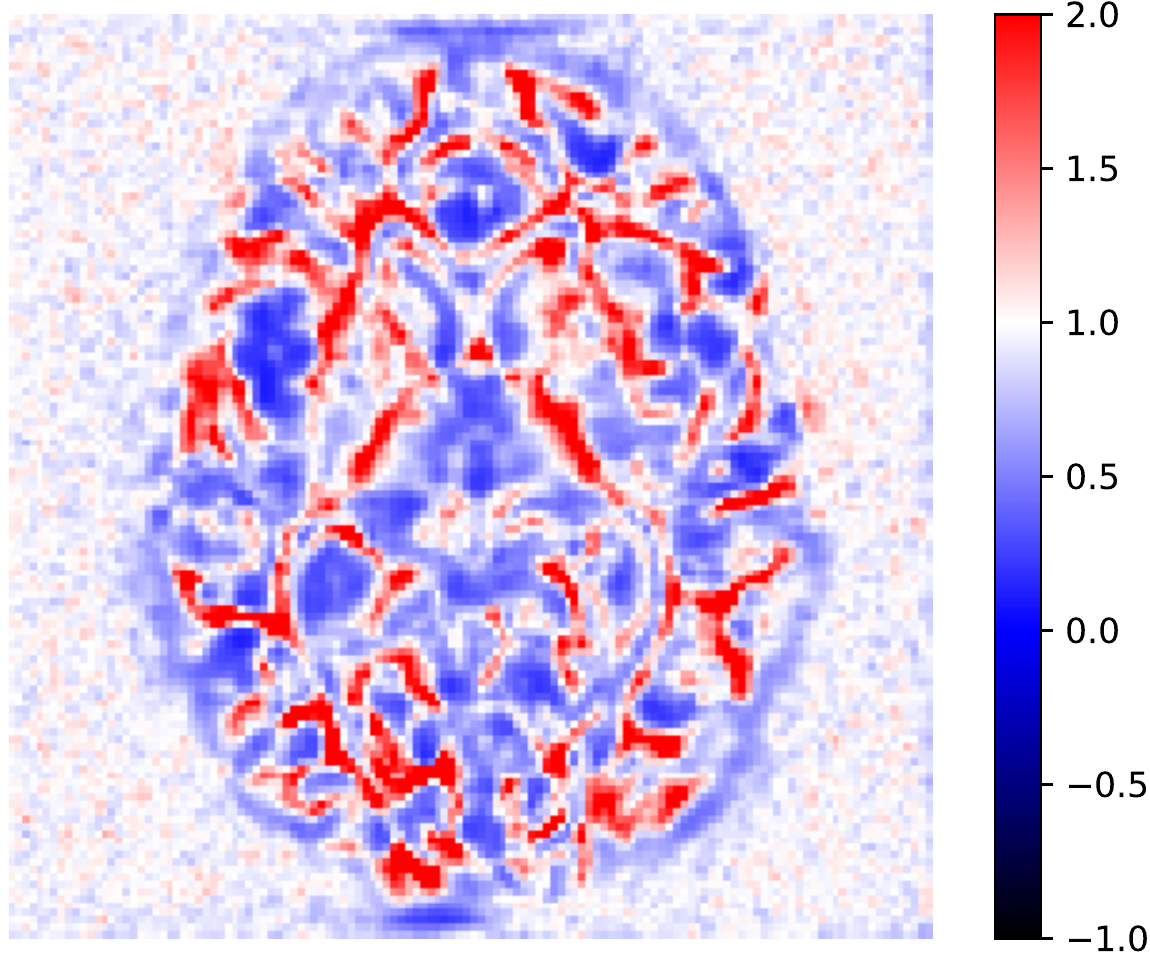}} 
         \end{tabular}
        &
        \begin{tabular}{c}
             \rotatebox{270}{\small{SyN}} \\ \\ \\ \\
             \rotatebox{270}{\small{VM}} \\ \\ \\ \\
             \rotatebox{270}{\small{Ours}}
         \end{tabular} 
    \end{tabular}
    \caption{Registration comparison among SyN, VM (VoxelMorph), and our DDR-Net. Left to right: the image pair median slice from OASIS dataset, the source image $I_0$ and the target image $I_1$; the warped image; the image difference between warped image and the target image; the deformation $\phi$; the determinant of the deformation Jacobian.}
    \label{fig:comparison_oasis3d}
\end{figure}

\begin{figure}[t]
    \centering
    \begin{tabular}{c|ccccl}
         Image Pair & $\phi \cdot I_0$ & $\phi \cdot I_0 - I_1$ &  $\phi$ &  $det(J_{\phi})$ \\
         \begin{tabular}{c}
            $I_0$ \\
            \raisebox{-.5\height}{\includegraphics[height=0.16\textwidth]{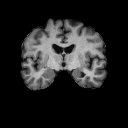}} \\ \\
            $I_1$ \\
            \raisebox{-.5\height}{\includegraphics[height=0.16\textwidth]{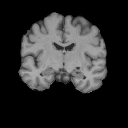}}
        \end{tabular}
       &
       \begin{tabular}{c}
             \raisebox{-.5\height}{\includegraphics[height=0.16\textwidth]{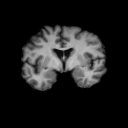}}\\ 
             \raisebox{-.5\height}{\includegraphics[height=0.16\textwidth]{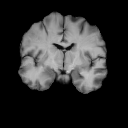}} \\
             \raisebox{-.5\height}{\includegraphics[height=0.16\textwidth]{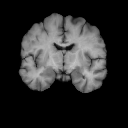}} 
       \end{tabular}
        &
        \begin{tabular}{c}
             \raisebox{-.5\height}{\includegraphics[height=0.16\textwidth]{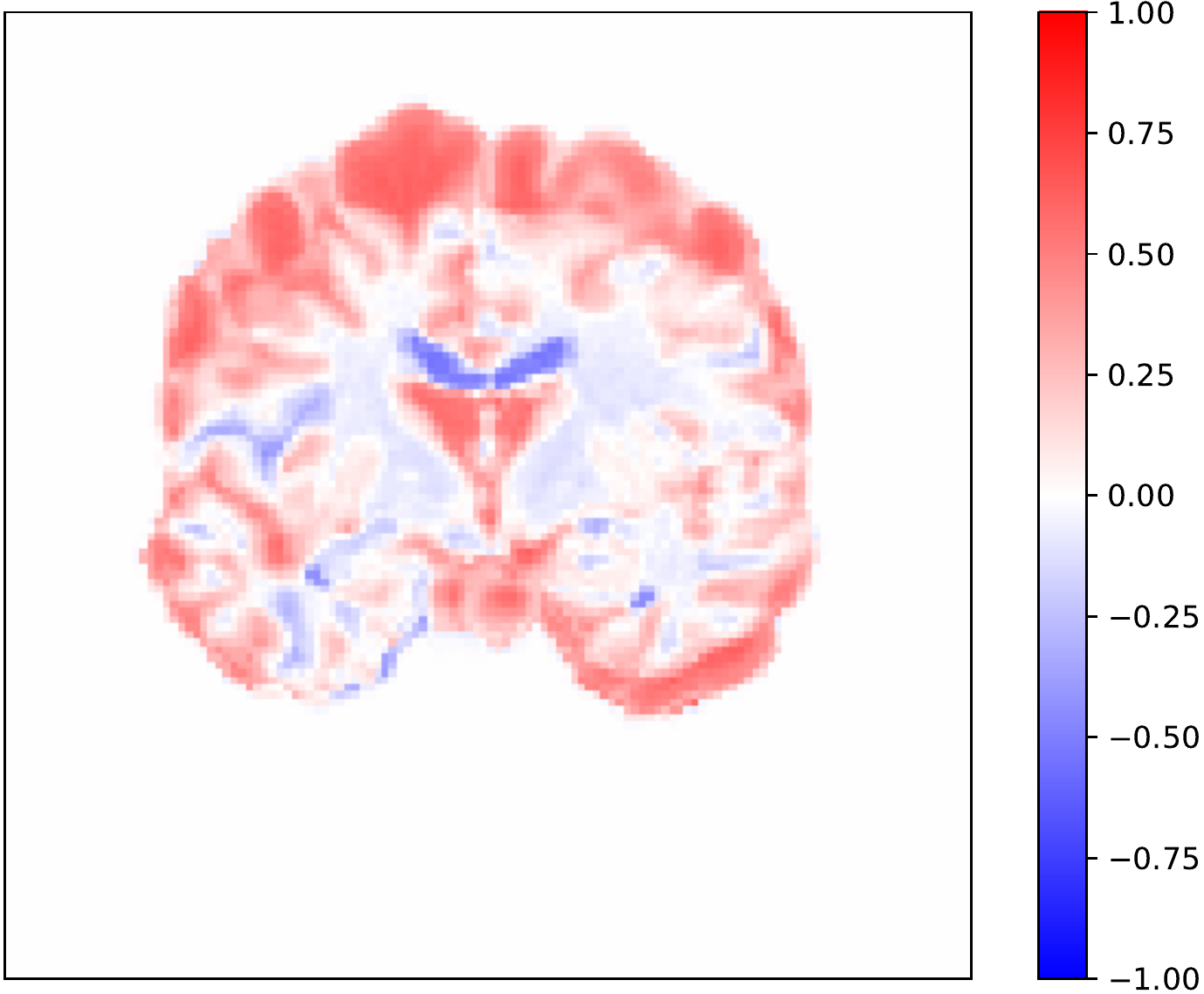}} \\ 
             \raisebox{-.5\height}{\includegraphics[height=0.16\textwidth]{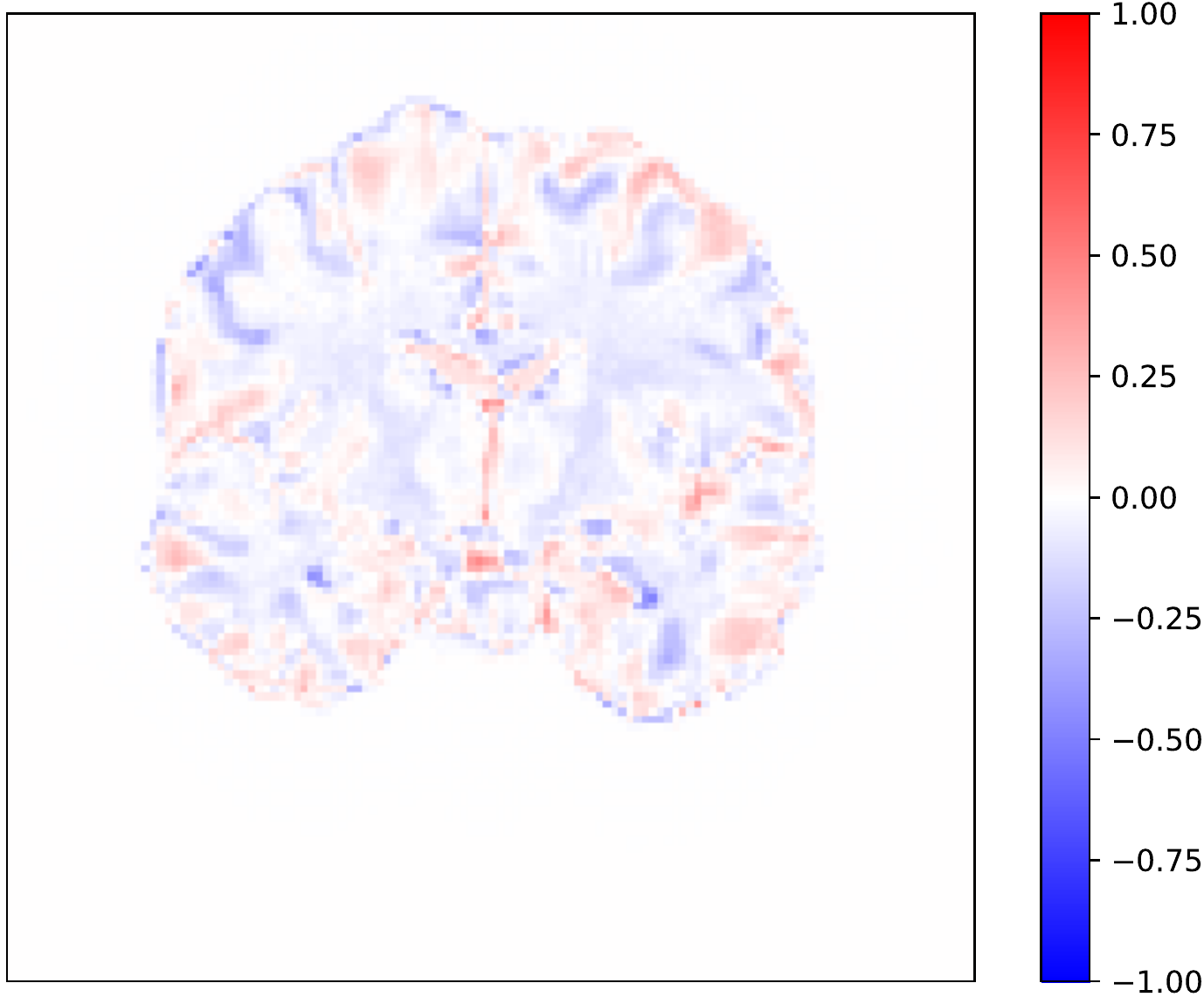}} \\
             \raisebox{-.5\height}{\includegraphics[height=0.16\textwidth]{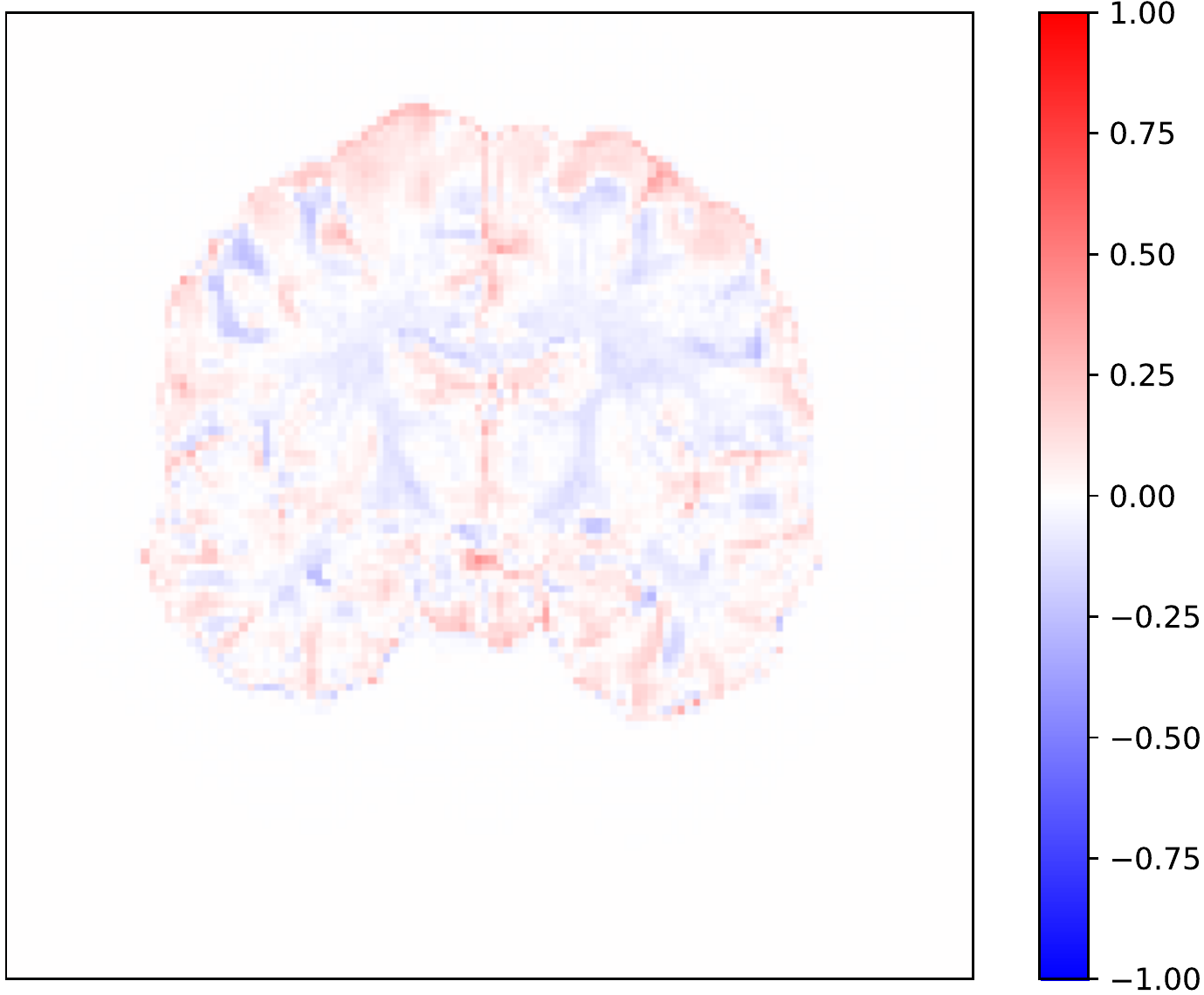}}  
         \end{tabular}
       &
        \begin{tabular}{c}
             \raisebox{-.5\height}{\includegraphics[height=0.16\textwidth]{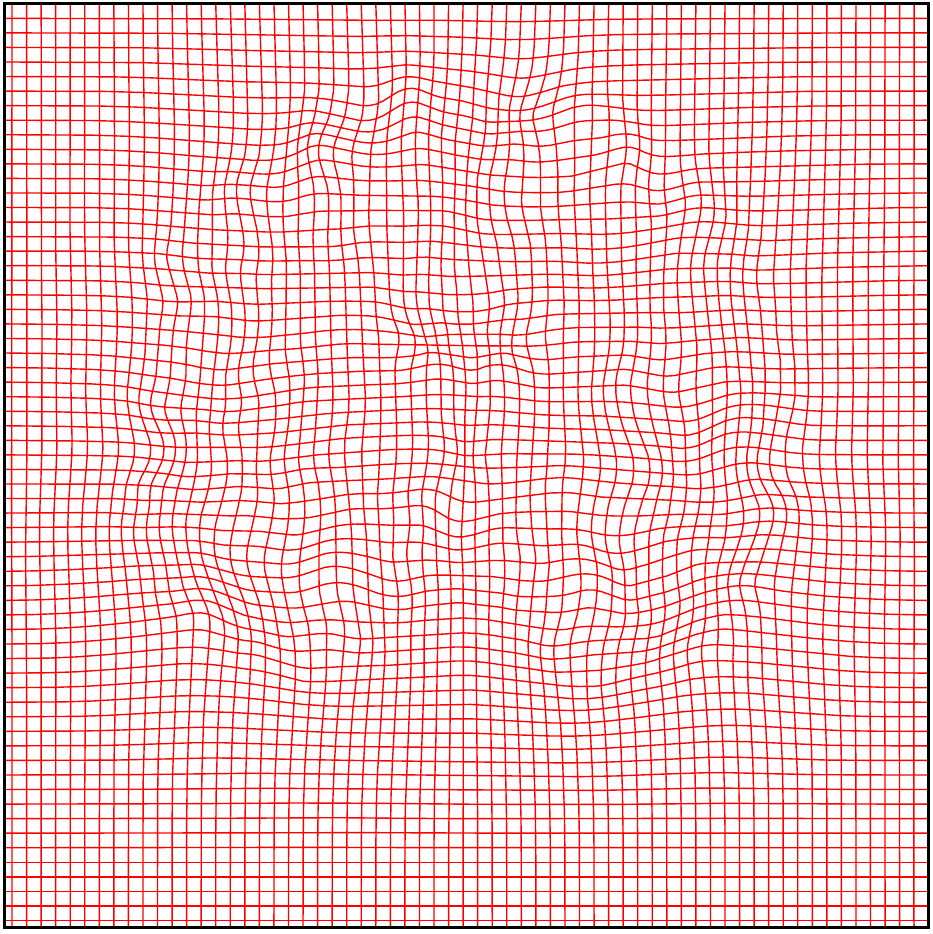}} \\ 
             \raisebox{-.5\height}{\includegraphics[height=0.16\textwidth]{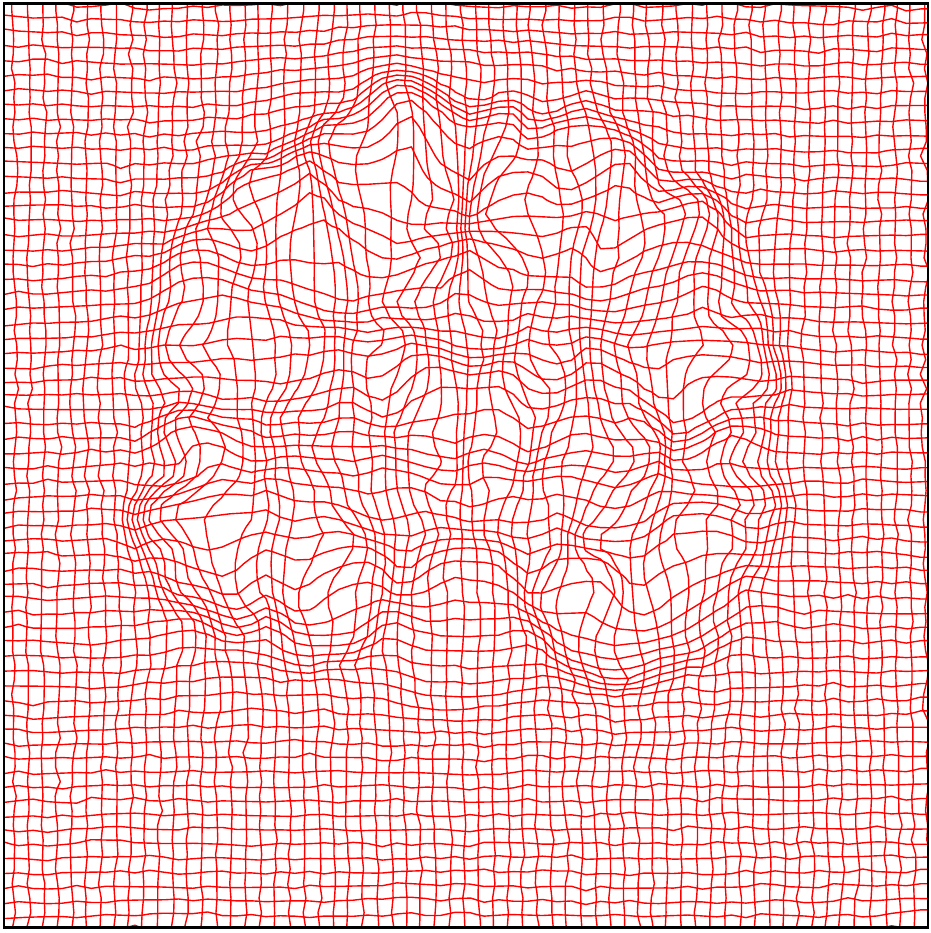}} \\
             \raisebox{-.5\height}{\includegraphics[height=0.16\textwidth]{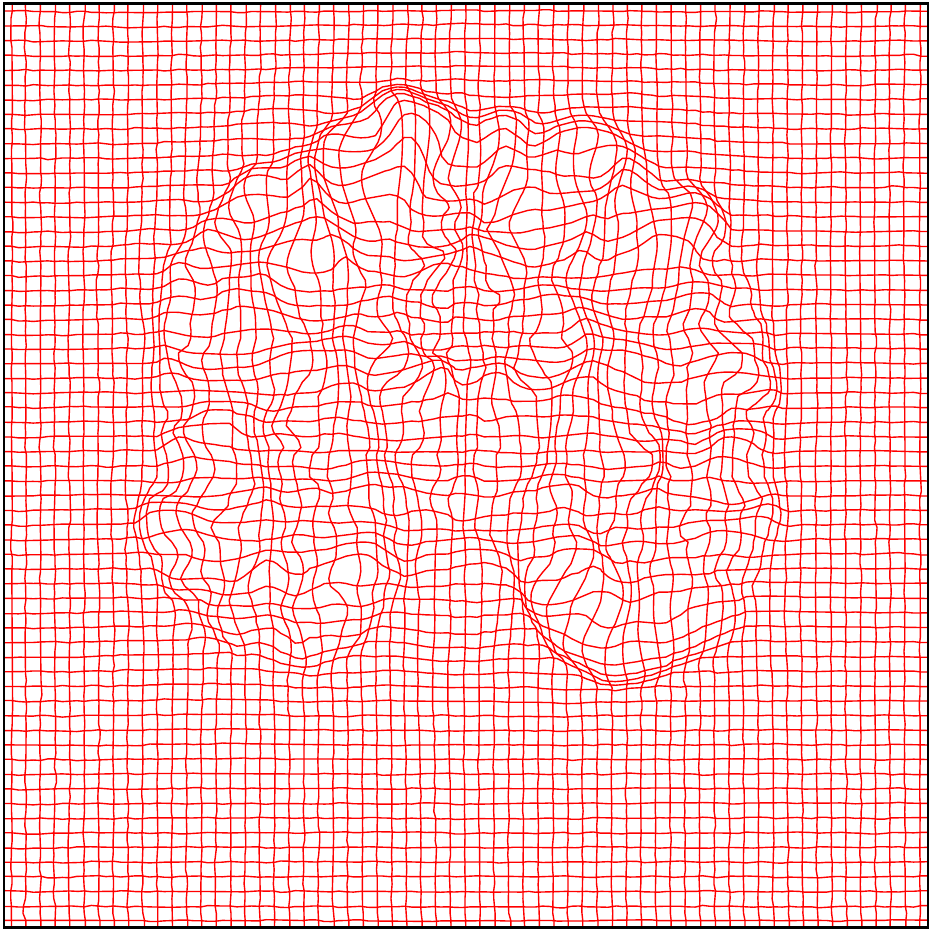}} 
         \end{tabular}
       
       &
       \begin{tabular}{c}
             \raisebox{-.5\height}{\includegraphics[height=0.16\textwidth]{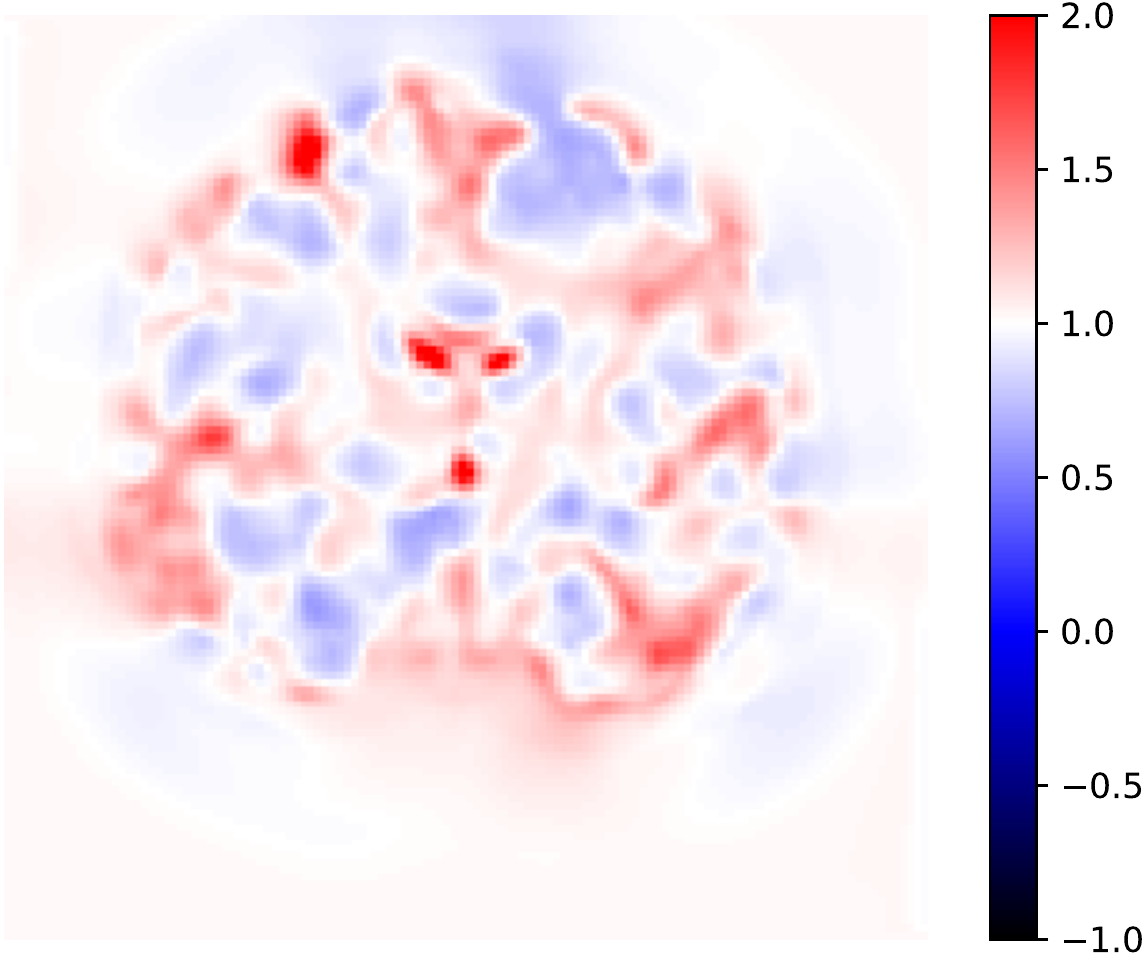}} \\ 
             \raisebox{-.5\height}{\includegraphics[height=0.16\textwidth]{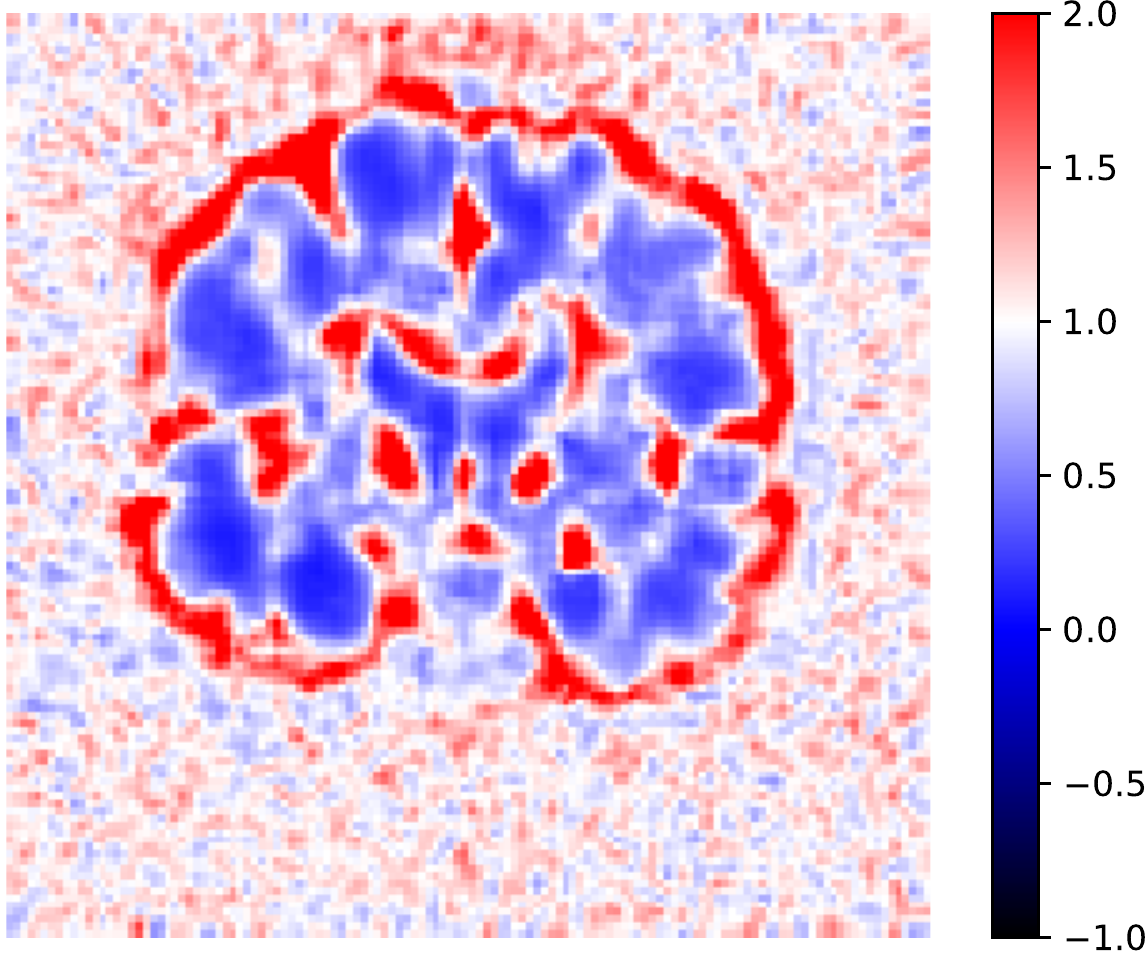}} \\
             \raisebox{-.5\height}{\includegraphics[height=0.16\textwidth]{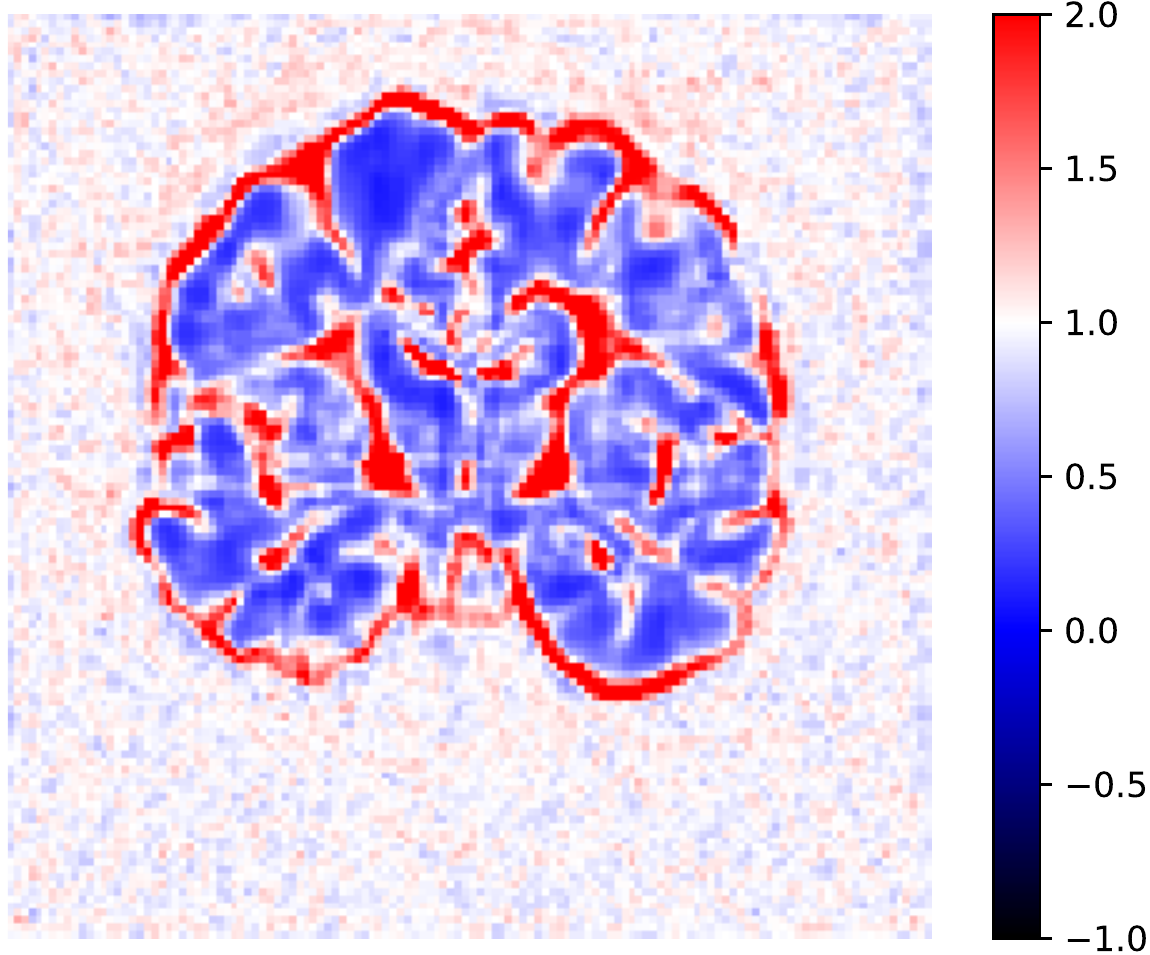}} 
         \end{tabular}
        &
        \begin{tabular}{c}
             \rotatebox{270}{\small{SyN}} \\ \\ \\ \\
             \rotatebox{270}{\small{VM}} \\ \\ \\ \\
             \rotatebox{270}{\small{Ours}}
         \end{tabular} 
    \end{tabular}
    \caption{Registration comparison among SyN, VM (VoxelMorph), and our DDR-Net for the IBSR dataset. Left to right: the same with Fig.~\ref{fig:comparison_oasis3d}.} 
    \label{fig:comparison_ibsr3d}
\end{figure}

\begin{figure}[t]
    \centering
    \begin{tabular}{c|ccccl}
         Image Pair & $\phi \cdot I_0$ & $\phi \cdot I_0 - I_1$ &  $\phi$ &  $det(J_{\phi})$ \\
         \begin{tabular}{c}
            $I_0$ \\
            \raisebox{-.5\height}{\includegraphics[height=0.16\textwidth]{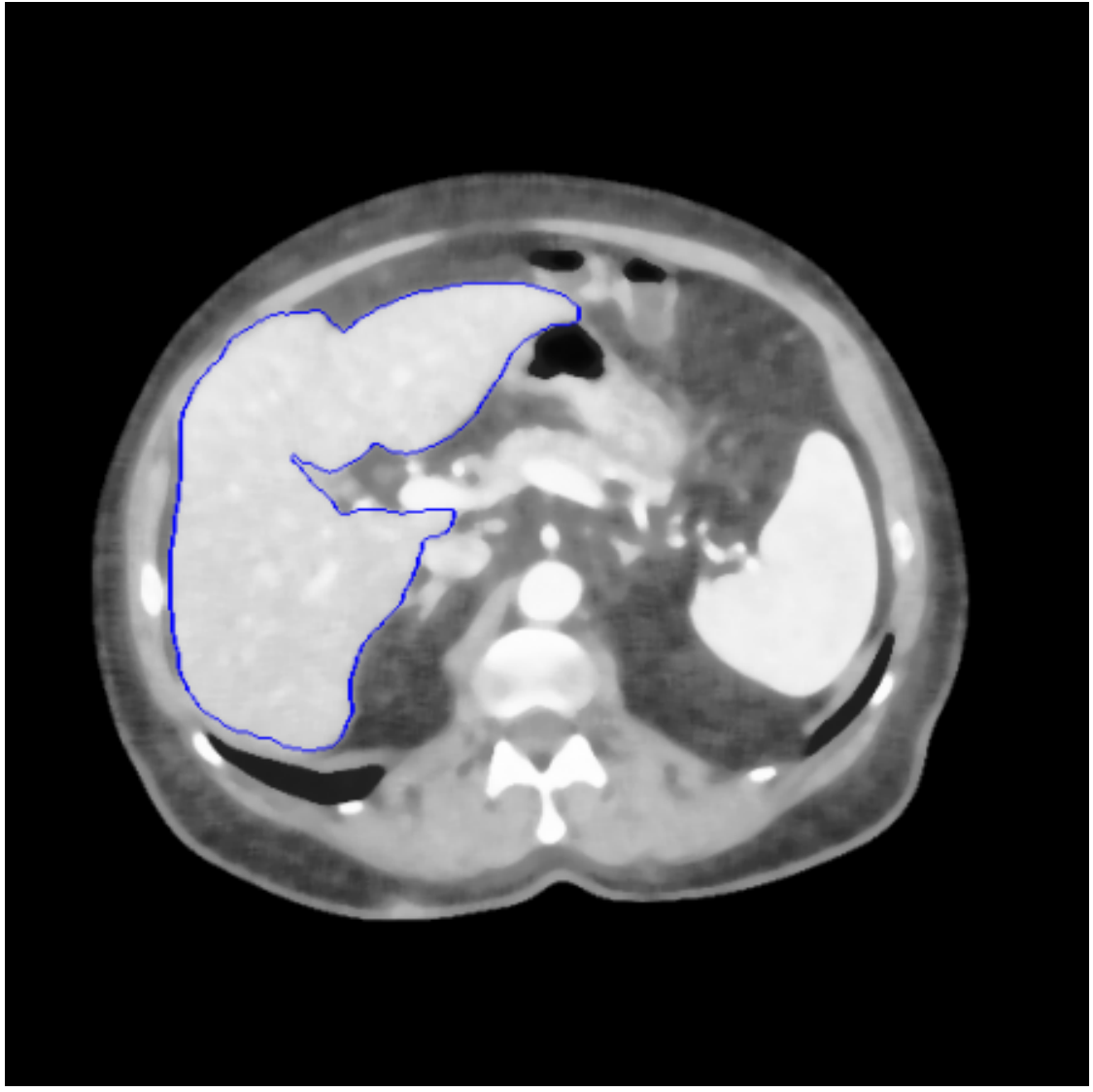}} \\ \\
            $I_1$ \\
            \raisebox{-.5\height}{\includegraphics[height=0.16\textwidth]{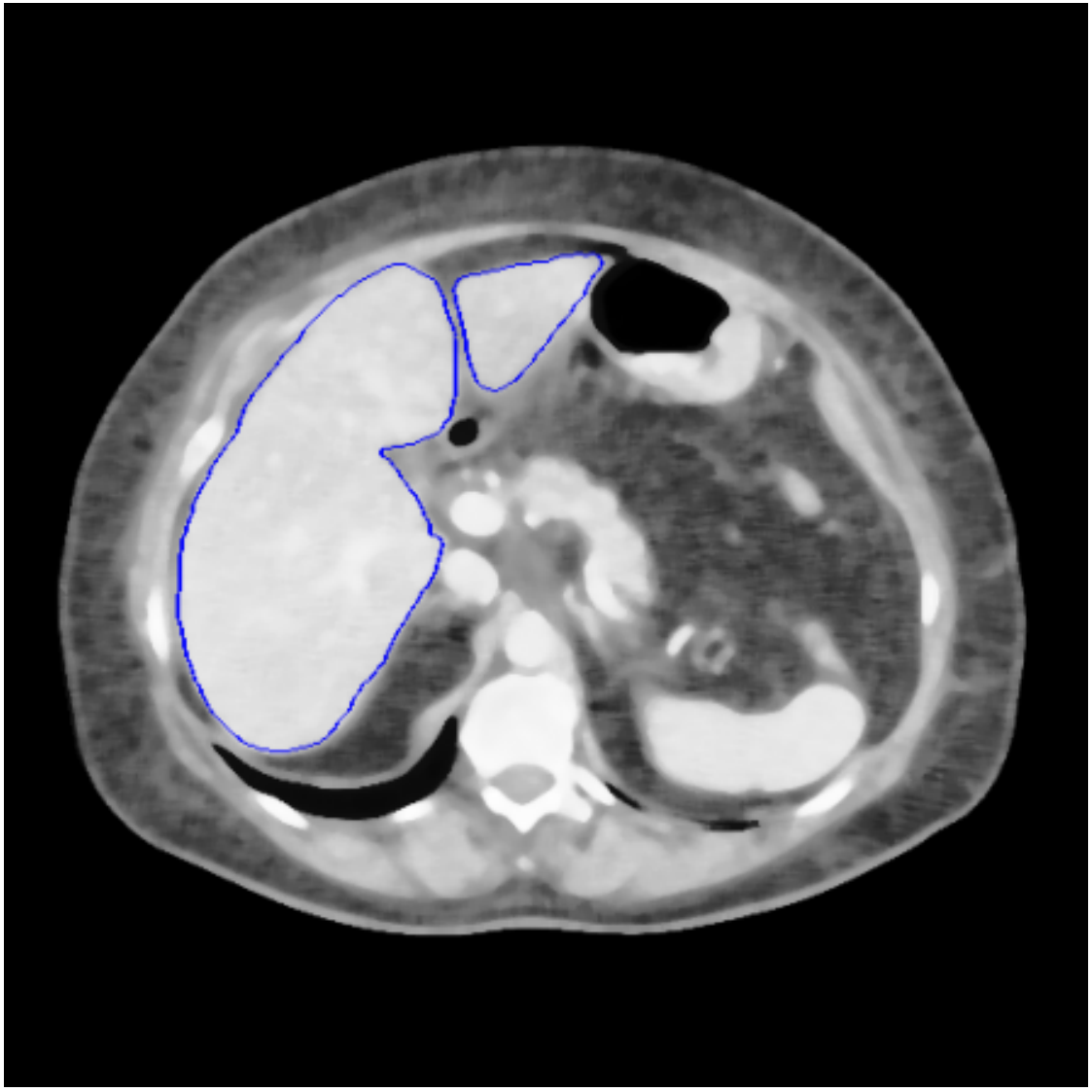}}
        \end{tabular}
       &
       \begin{tabular}{c}
             \raisebox{-.5\height}{\includegraphics[height=0.16\textwidth]{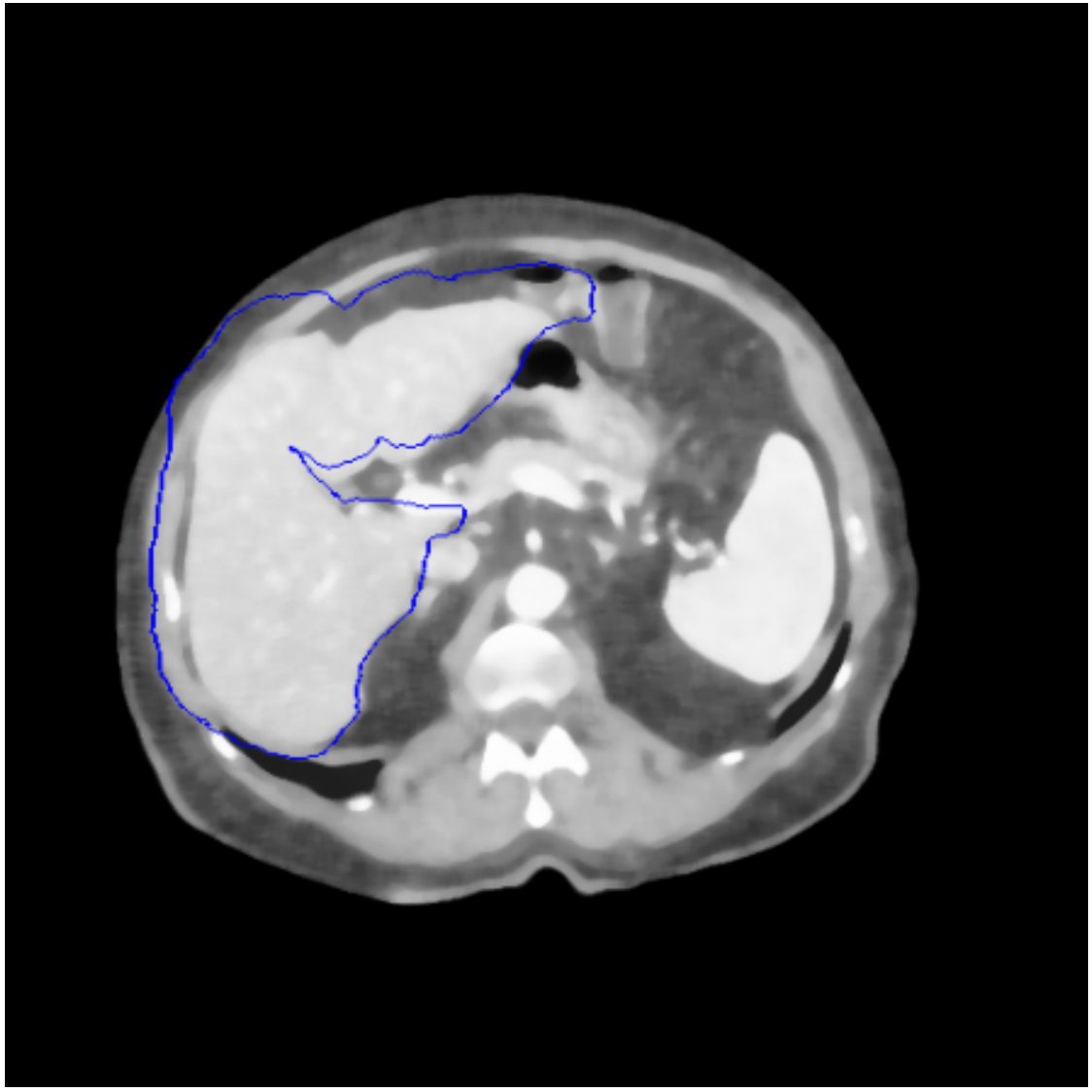}}\\ 
             \raisebox{-.5\height}{\includegraphics[height=0.16\textwidth]{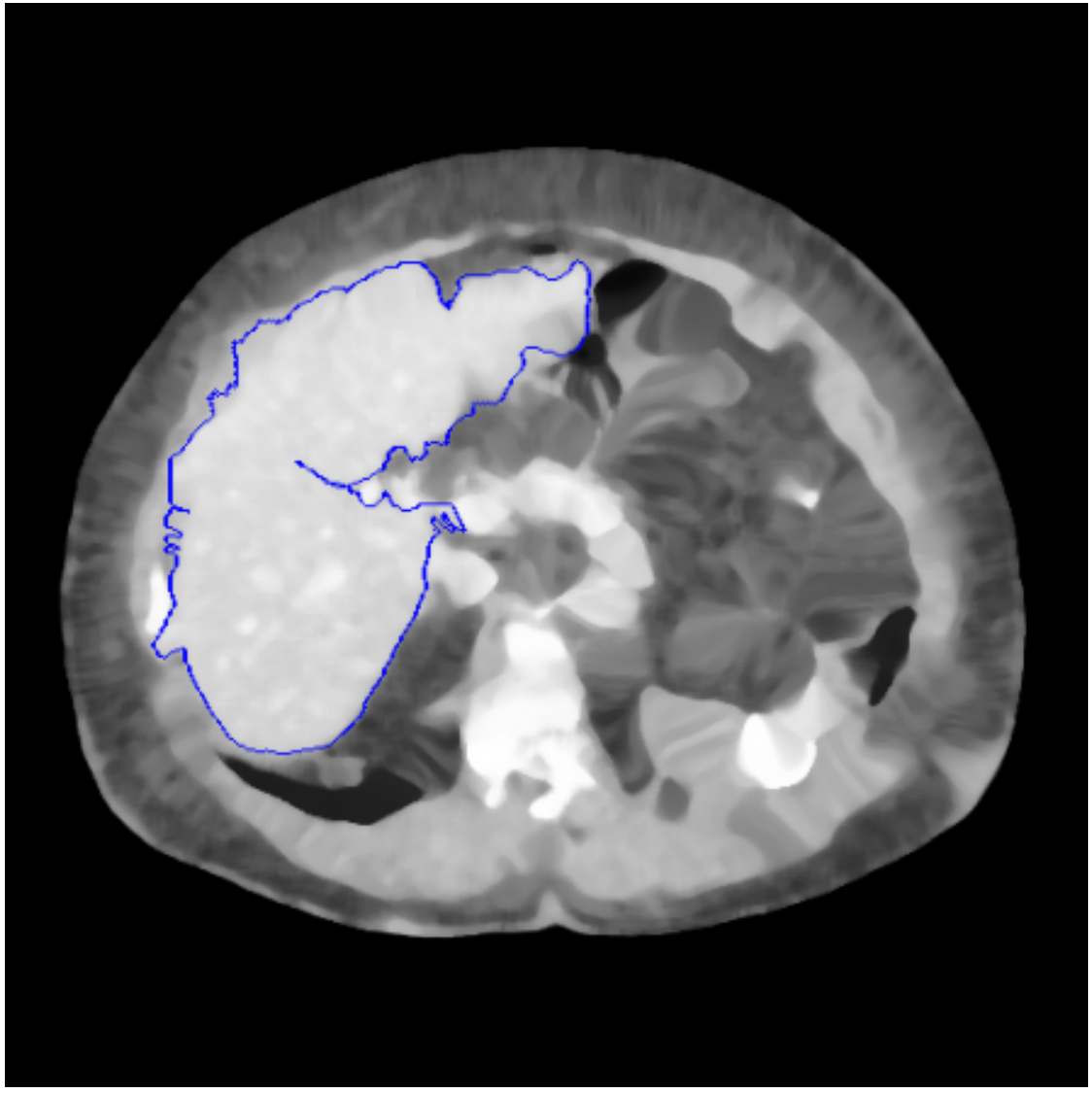}} \\
             \raisebox{-.5\height}{\includegraphics[height=0.16\textwidth]{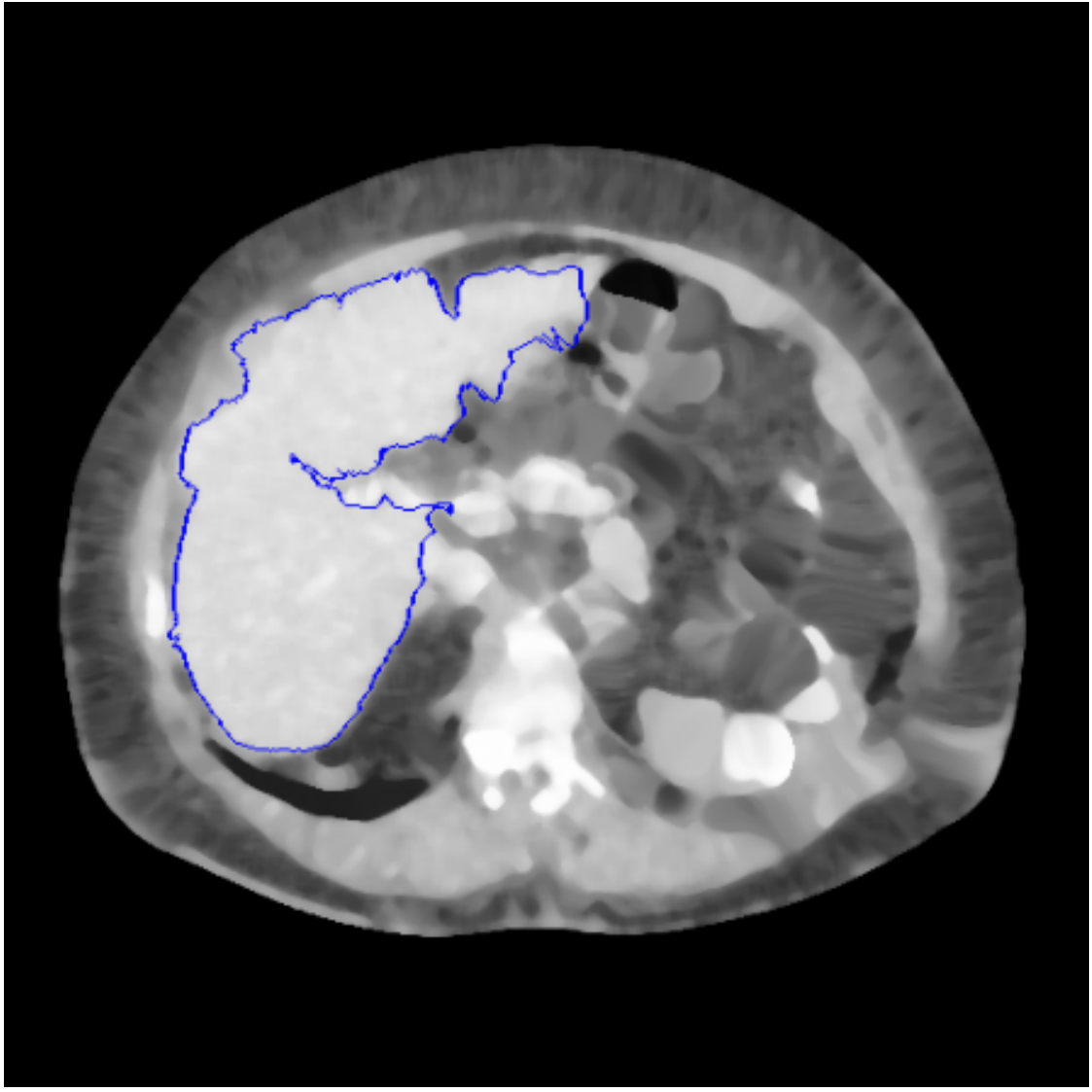}} 
       \end{tabular}
        &
        \begin{tabular}{c}
             \raisebox{-.5\height}{\includegraphics[height=0.16\textwidth]{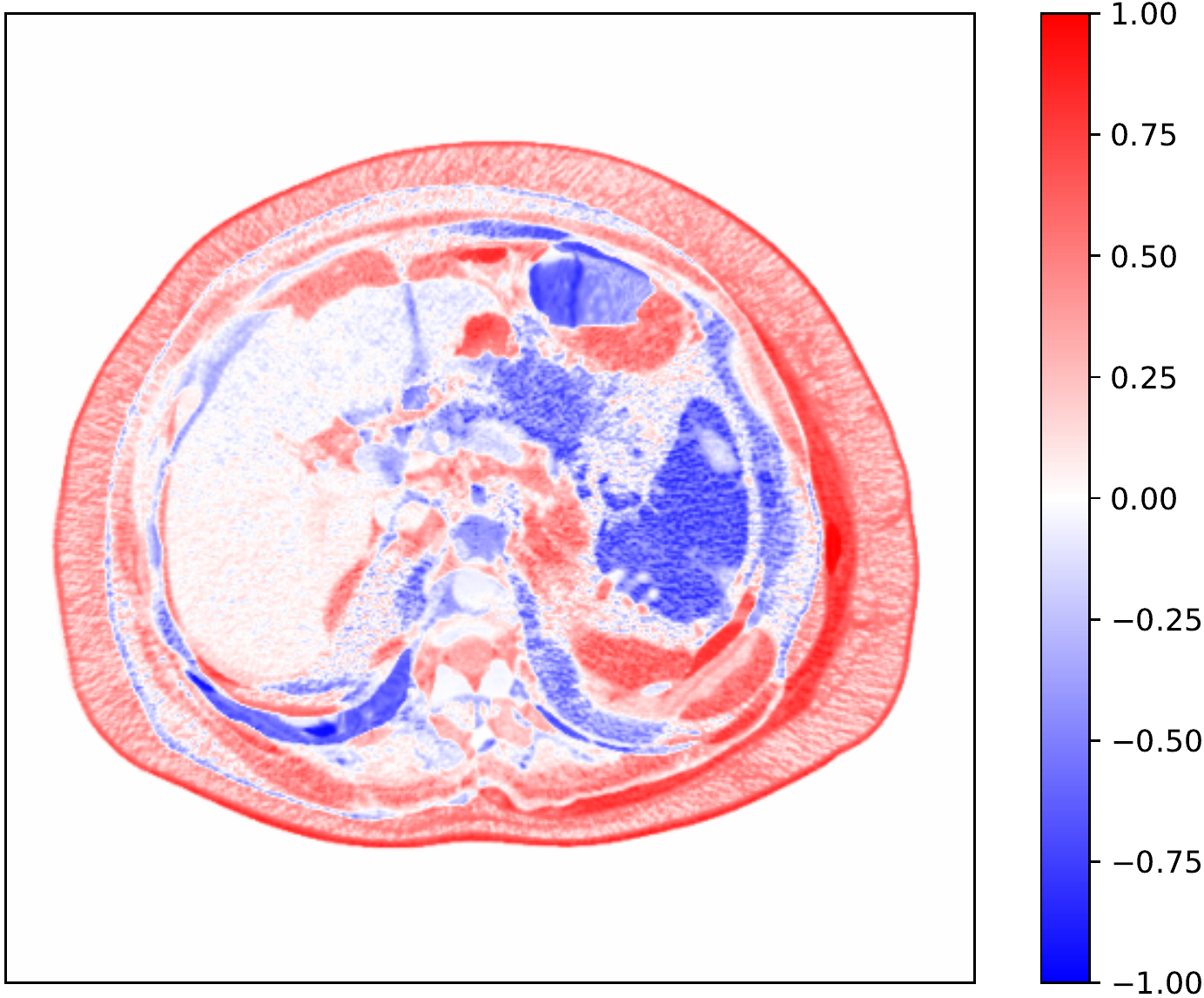}} \\ 
             \raisebox{-.5\height}{\includegraphics[height=0.16\textwidth]{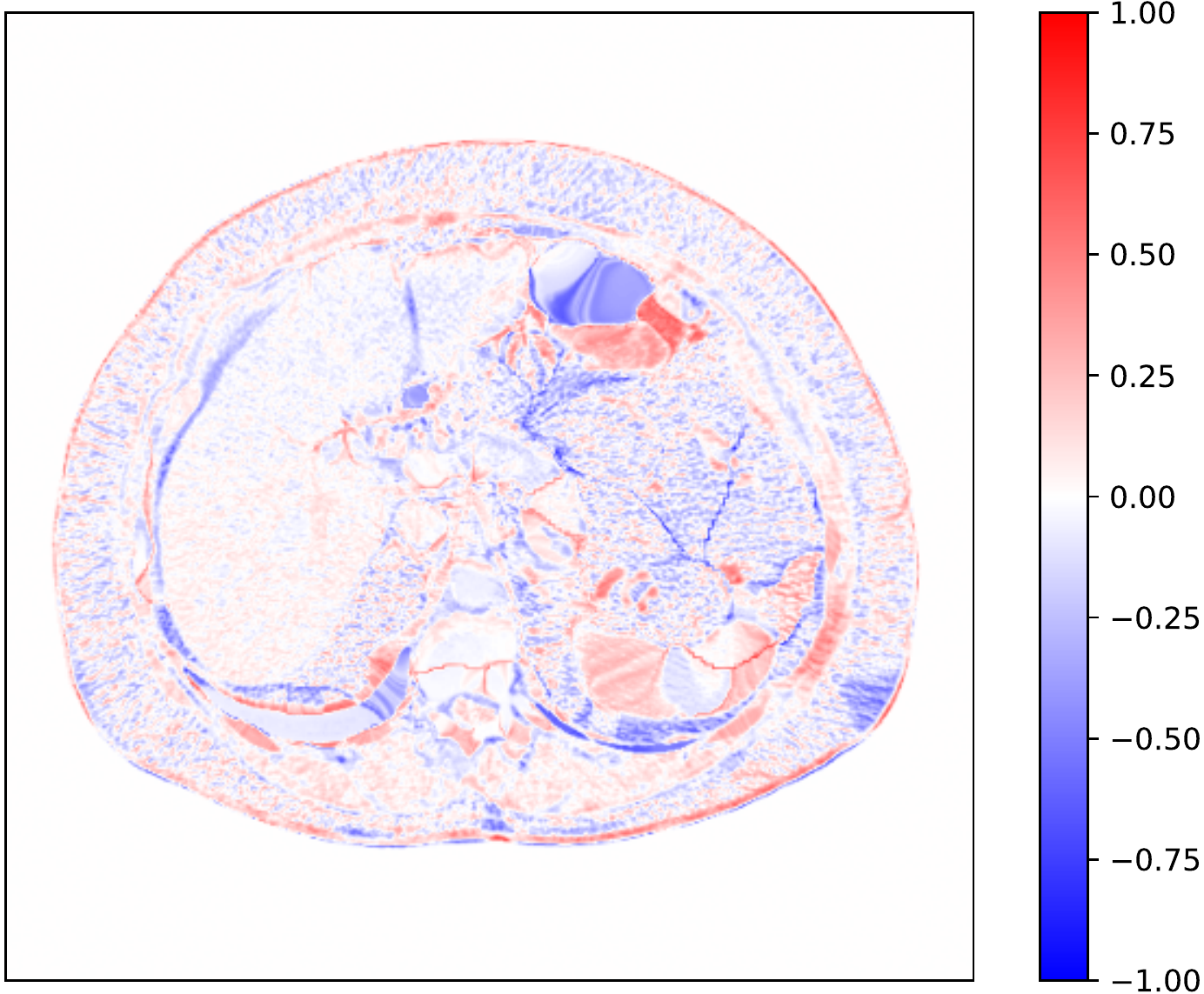}} \\
             \raisebox{-.5\height}{\includegraphics[height=0.16\textwidth]{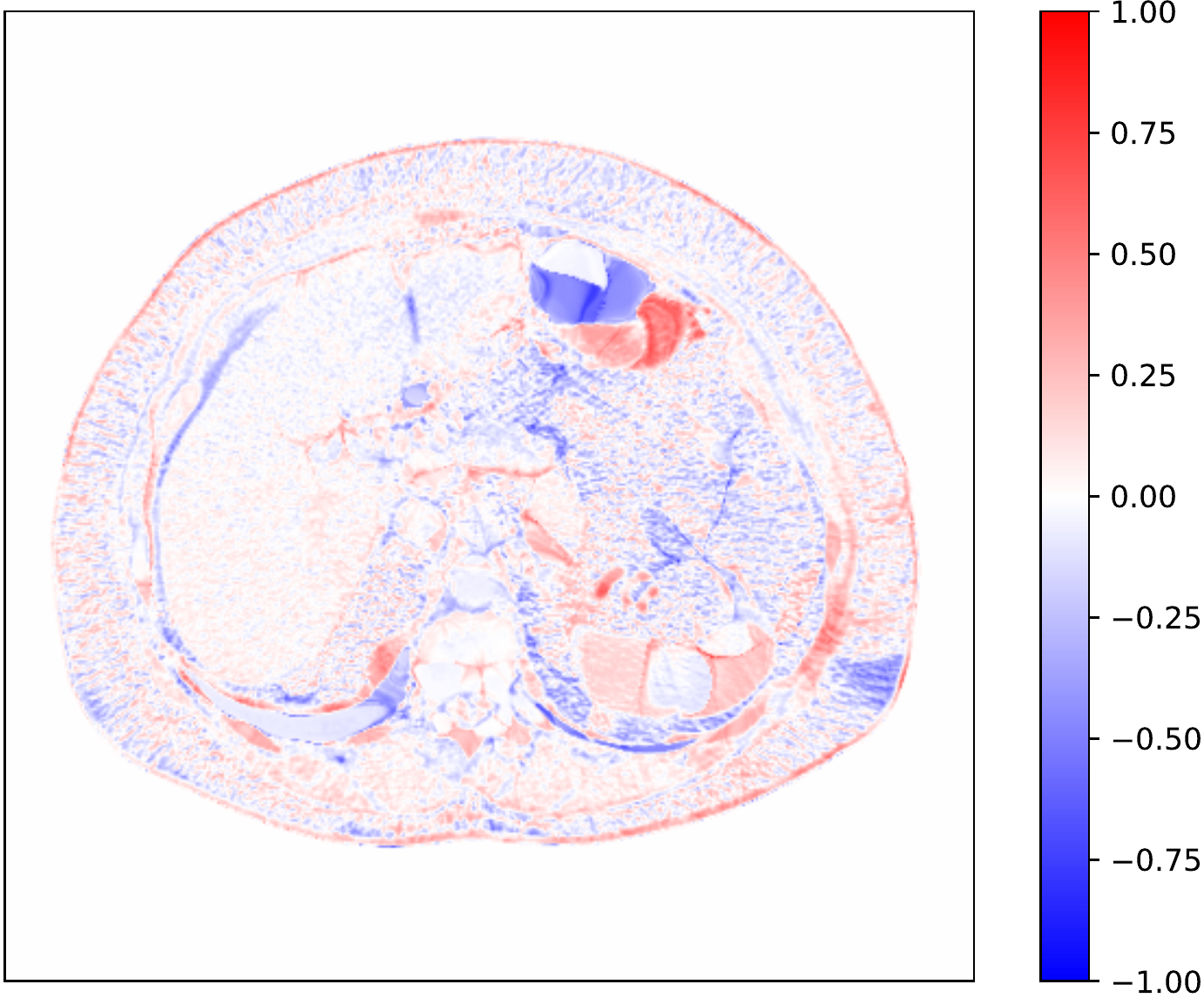}}  
         \end{tabular}
       &
        \begin{tabular}{c}
             \raisebox{-.5\height}{\includegraphics[height=0.16\textwidth]{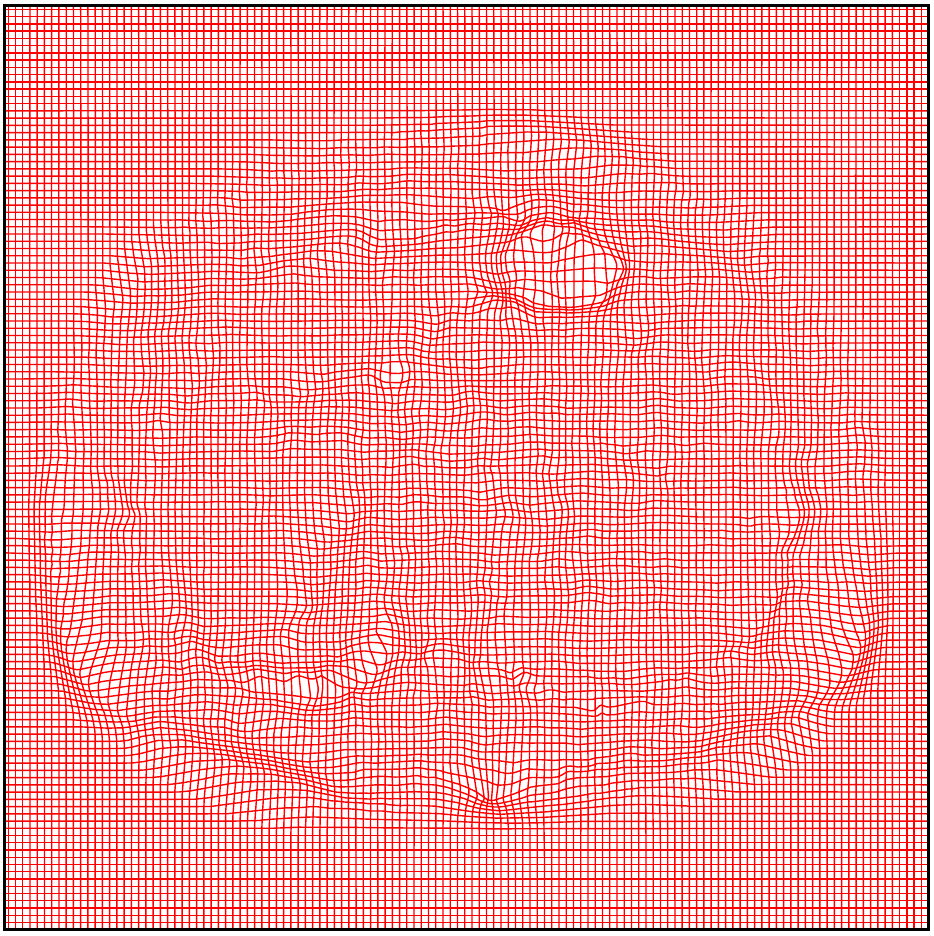}} \\ 
             \raisebox{-.5\height}{\includegraphics[height=0.16\textwidth]{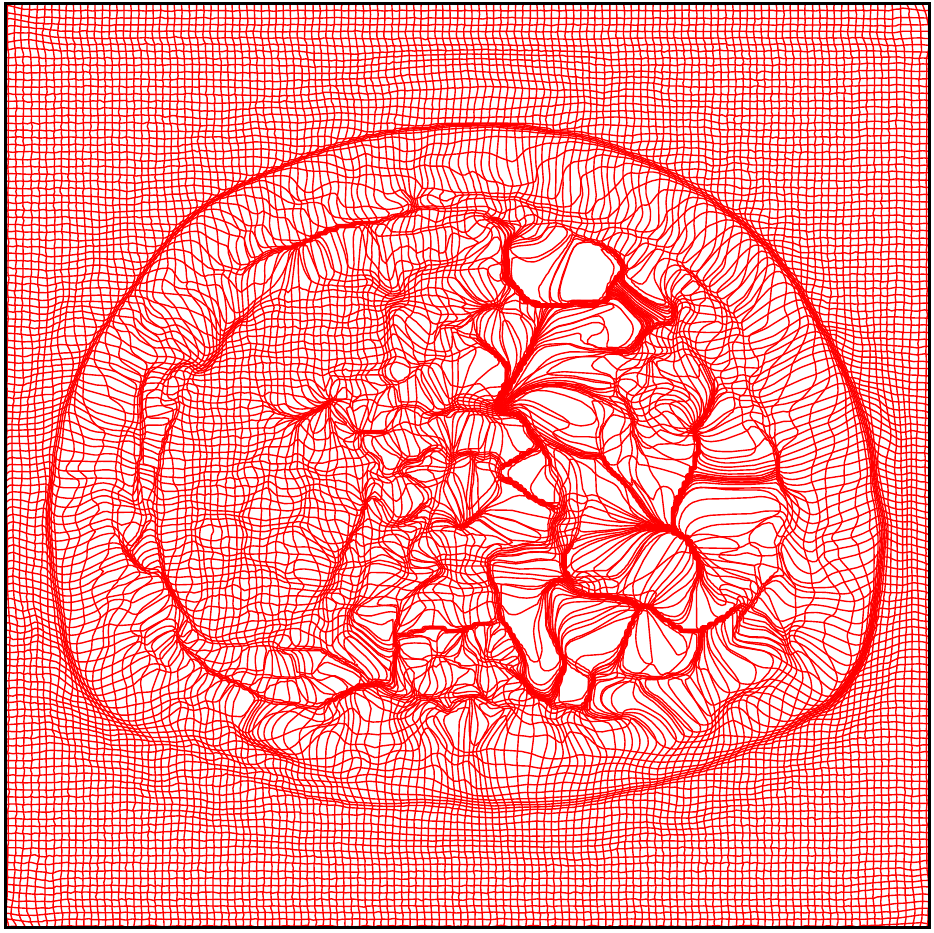}} \\
             \raisebox{-.5\height}{\includegraphics[height=0.16\textwidth]{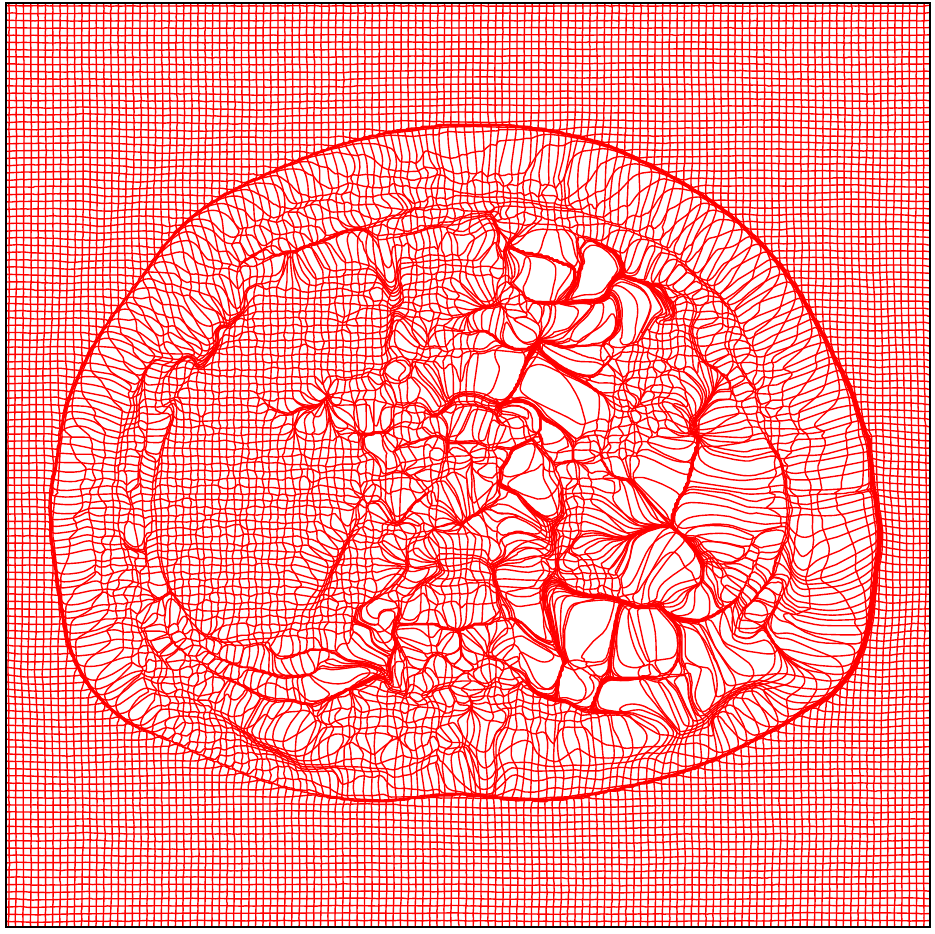}} 
         \end{tabular}
       
       &
       \begin{tabular}{c}
             \raisebox{-.5\height}{\includegraphics[height=0.16\textwidth]{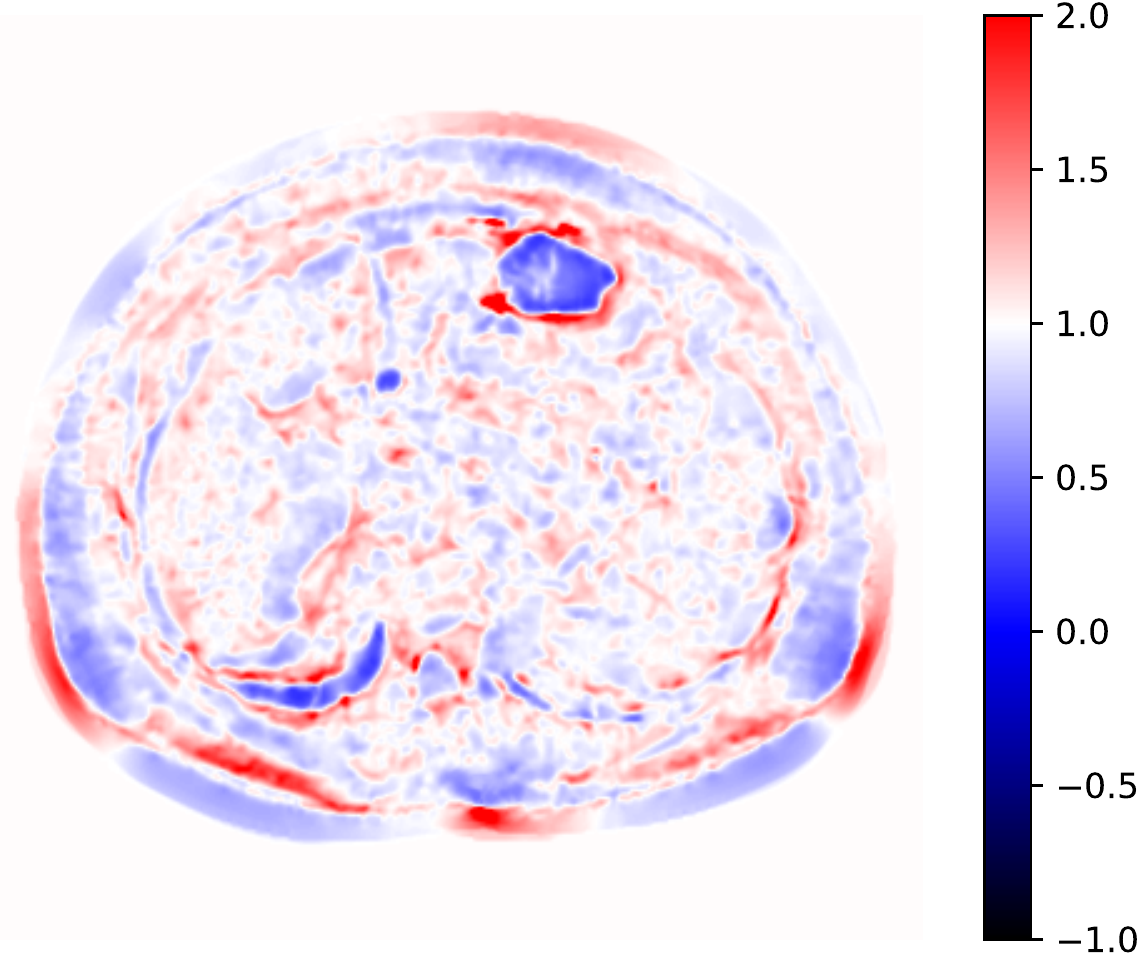}} \\ 
             \raisebox{-.5\height}{\includegraphics[height=0.16\textwidth]{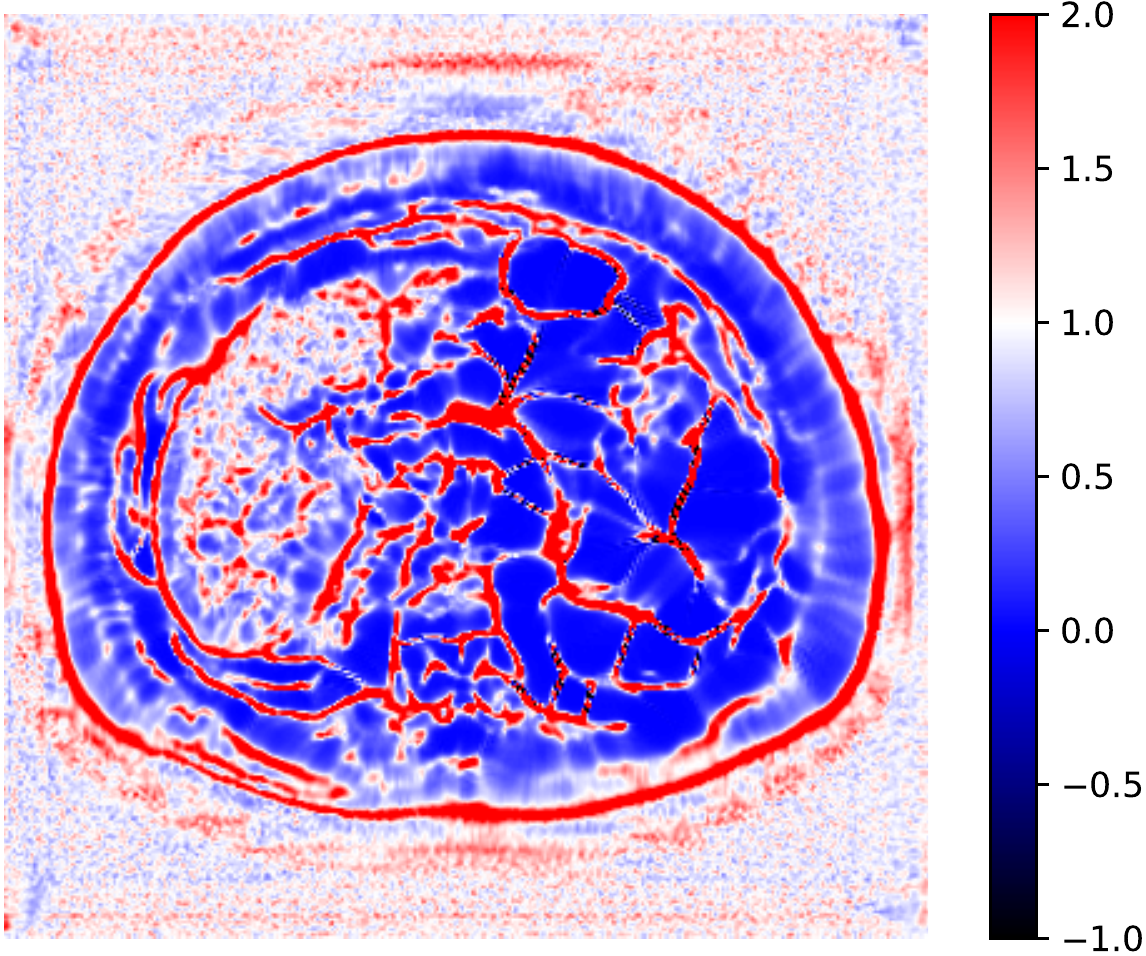}} \\
             \raisebox{-.5\height}{\includegraphics[height=0.16\textwidth]{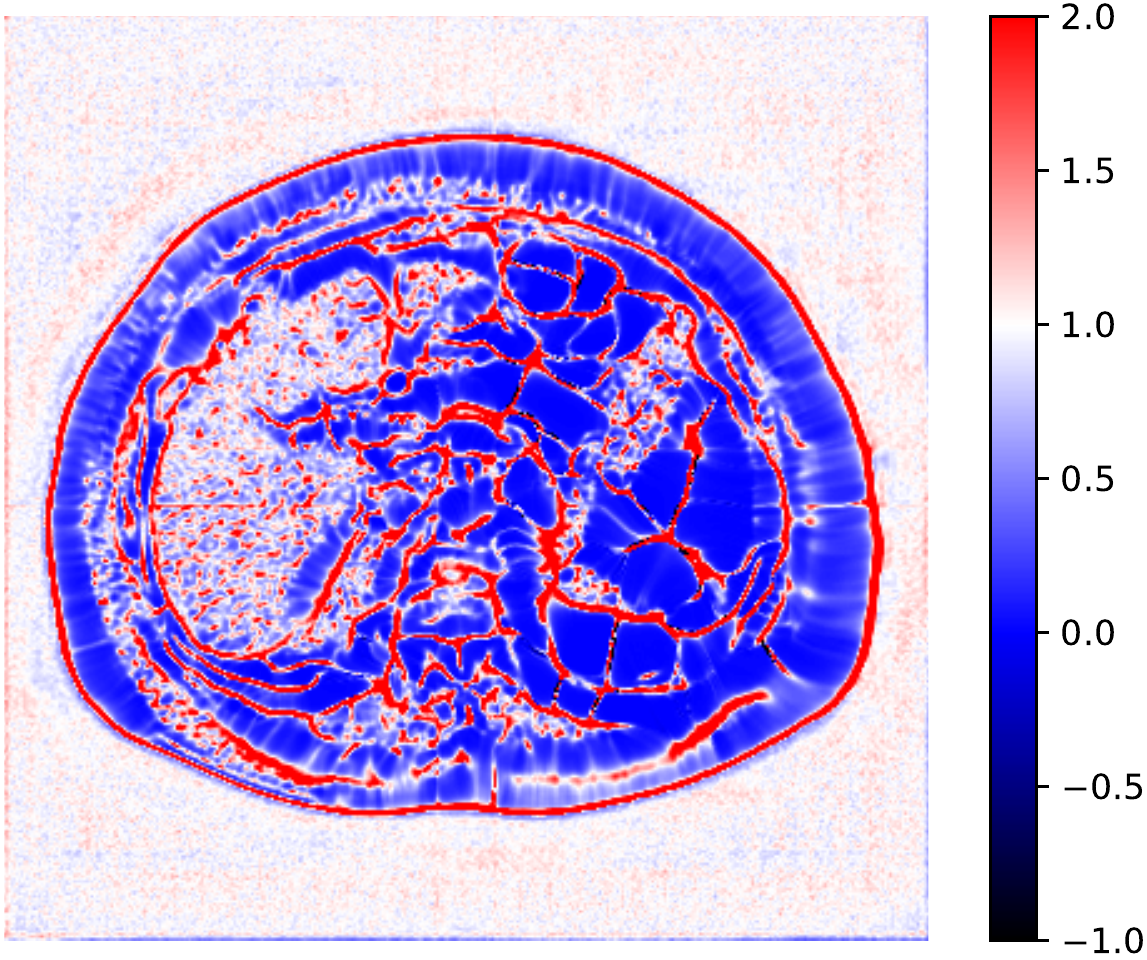}} 
         \end{tabular}
        &
        \begin{tabular}{c}
             \rotatebox{270}{\small{SyN}} \\ \\ \\ \\
             \rotatebox{270}{\small{VM}} \\ \\ \\ \\
             \rotatebox{270}{\small{Ours}}
         \end{tabular} \\
        
    \end{tabular}
    \caption{Registration comparison among SyN, VM (VoxelMorph), and our DDR-Net for the 3DIRCADB-01 dataset. Left to right: the same with Fig.~\ref{fig:comparison_oasis3d}.}
    \label{fig:comparison_3dircadb}
\end{figure}

\begin{figure}[t]
    \centering
    \begin{tabular}{ccccc}
         $I_0$ & $I_1$ & $\phi \cdot I_0$ (SyN) & $\phi \cdot I_0$ (VM) & $\phi \cdot I_0$ (Ours) \\
                 \raisebox{-.5\height}{\includegraphics[height=0.16\textwidth]{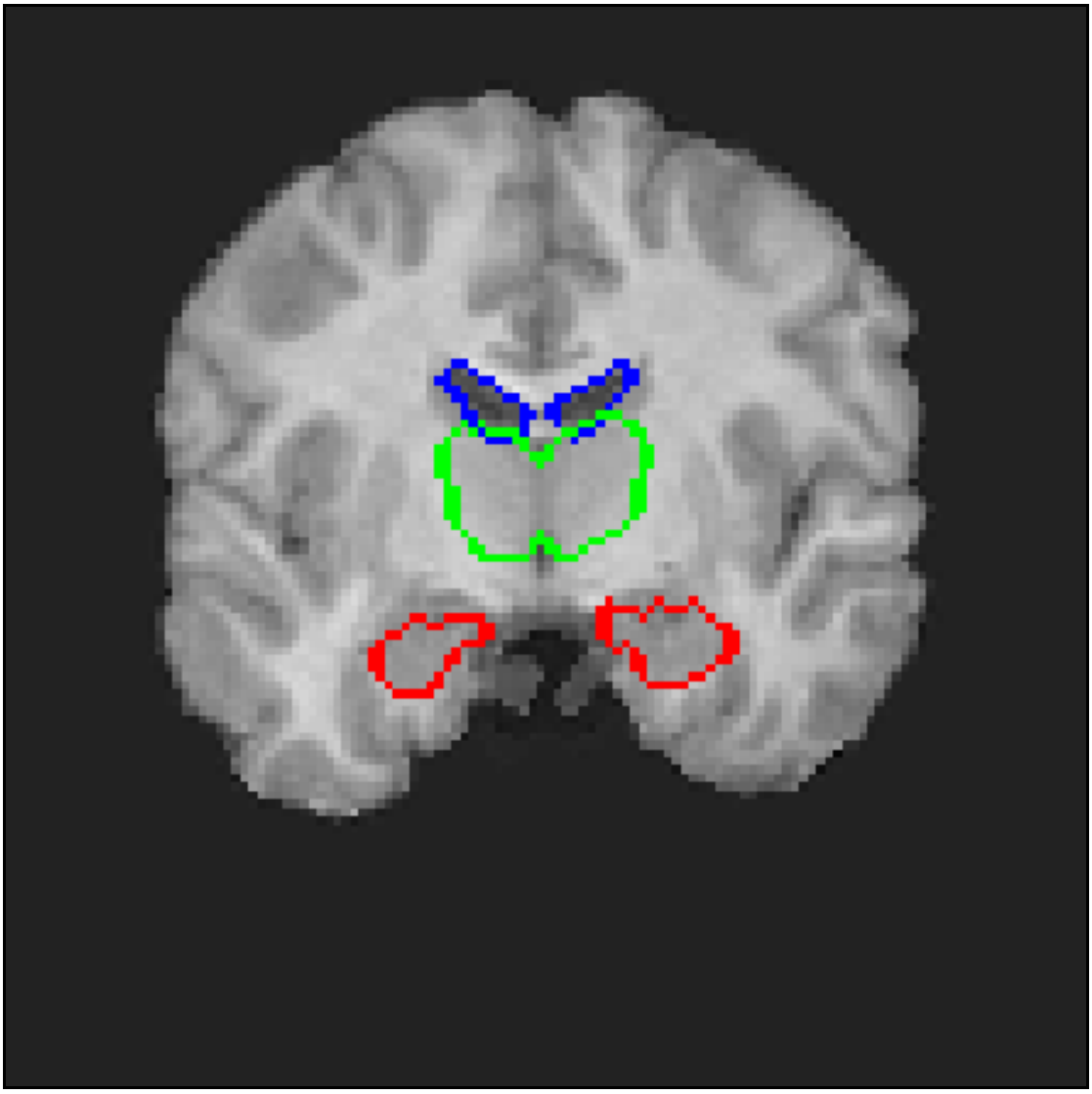}} &
                 \raisebox{-.5\height}{\includegraphics[height=0.16\textwidth]{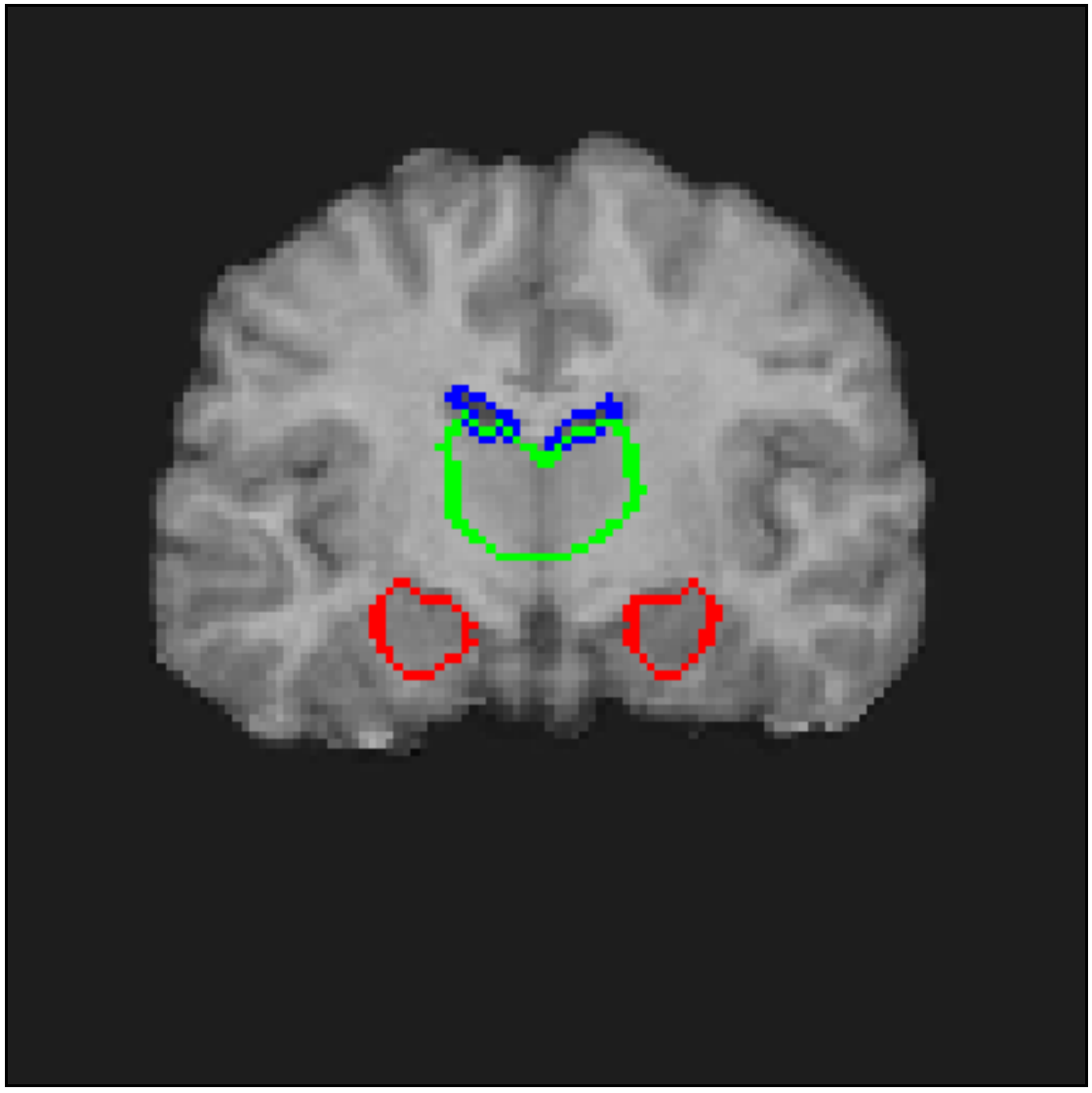}} &
                 \raisebox{-.5\height}{\includegraphics[height=0.16\textwidth]{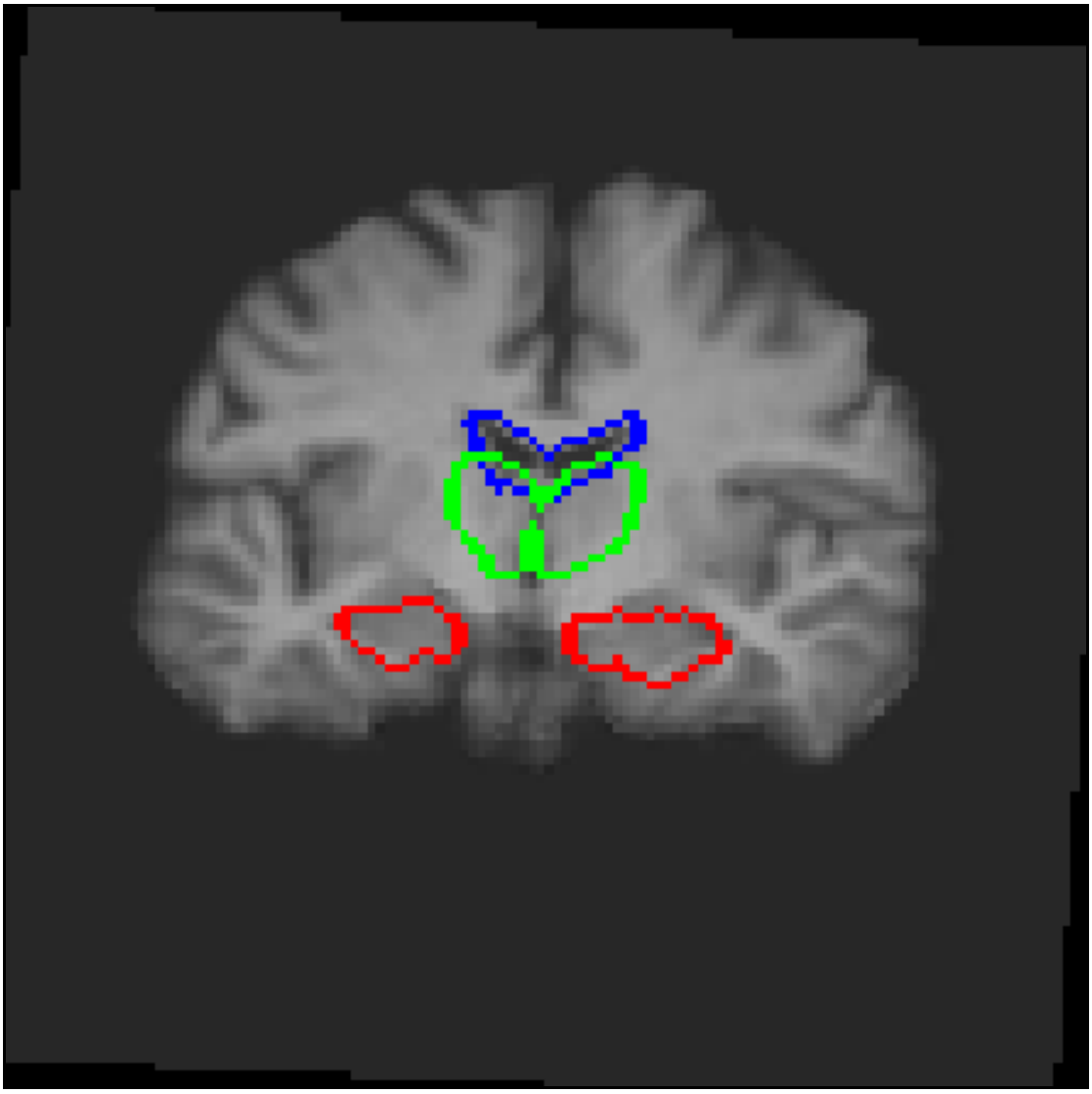}} &
                 \raisebox{-.5\height}{\includegraphics[height=0.16\textwidth]{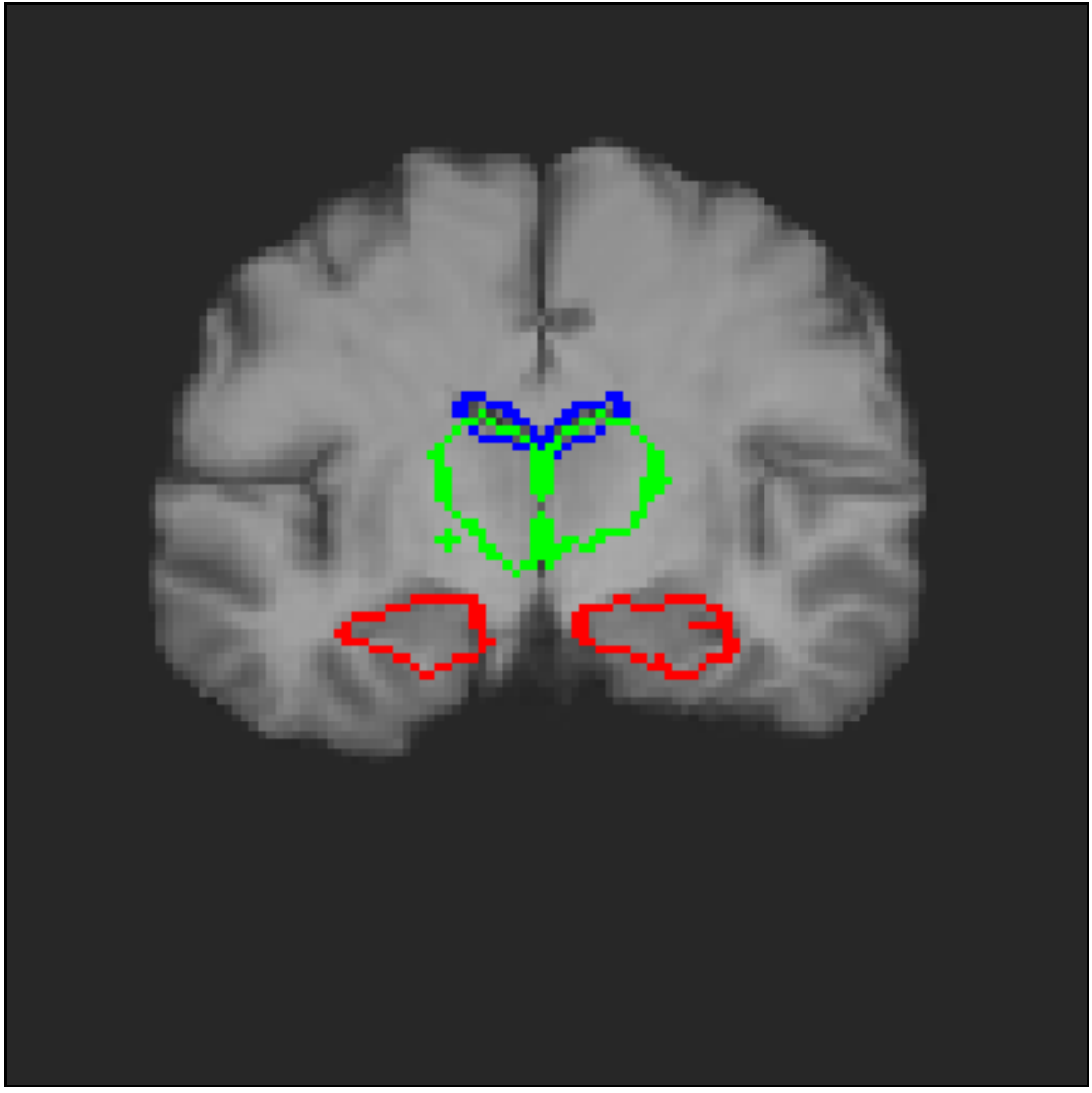}} &
                 \raisebox{-.5\height}{\includegraphics[height=0.16\textwidth]{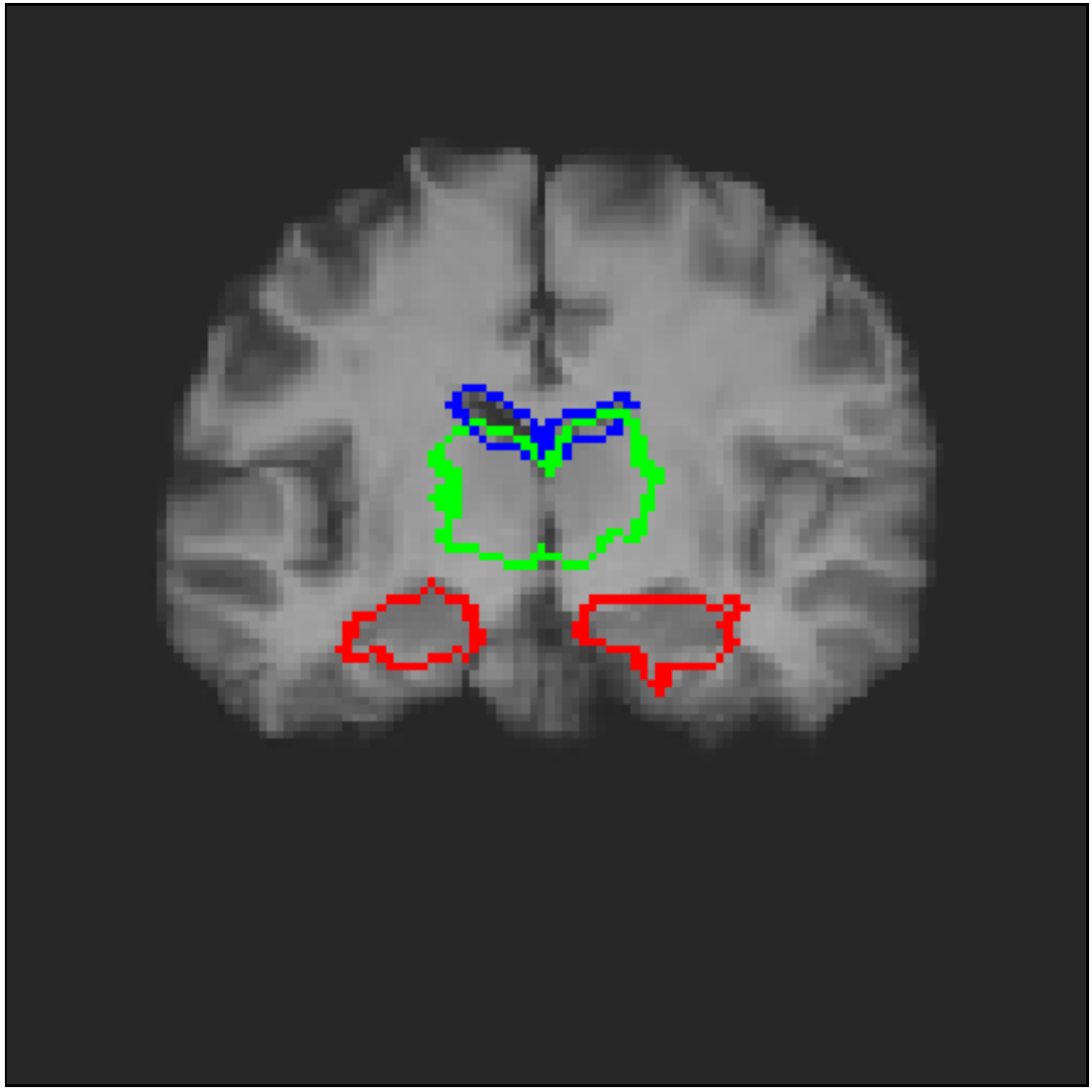}}\\
                 \raisebox{-.5\height}{\includegraphics[height=0.16\textwidth]{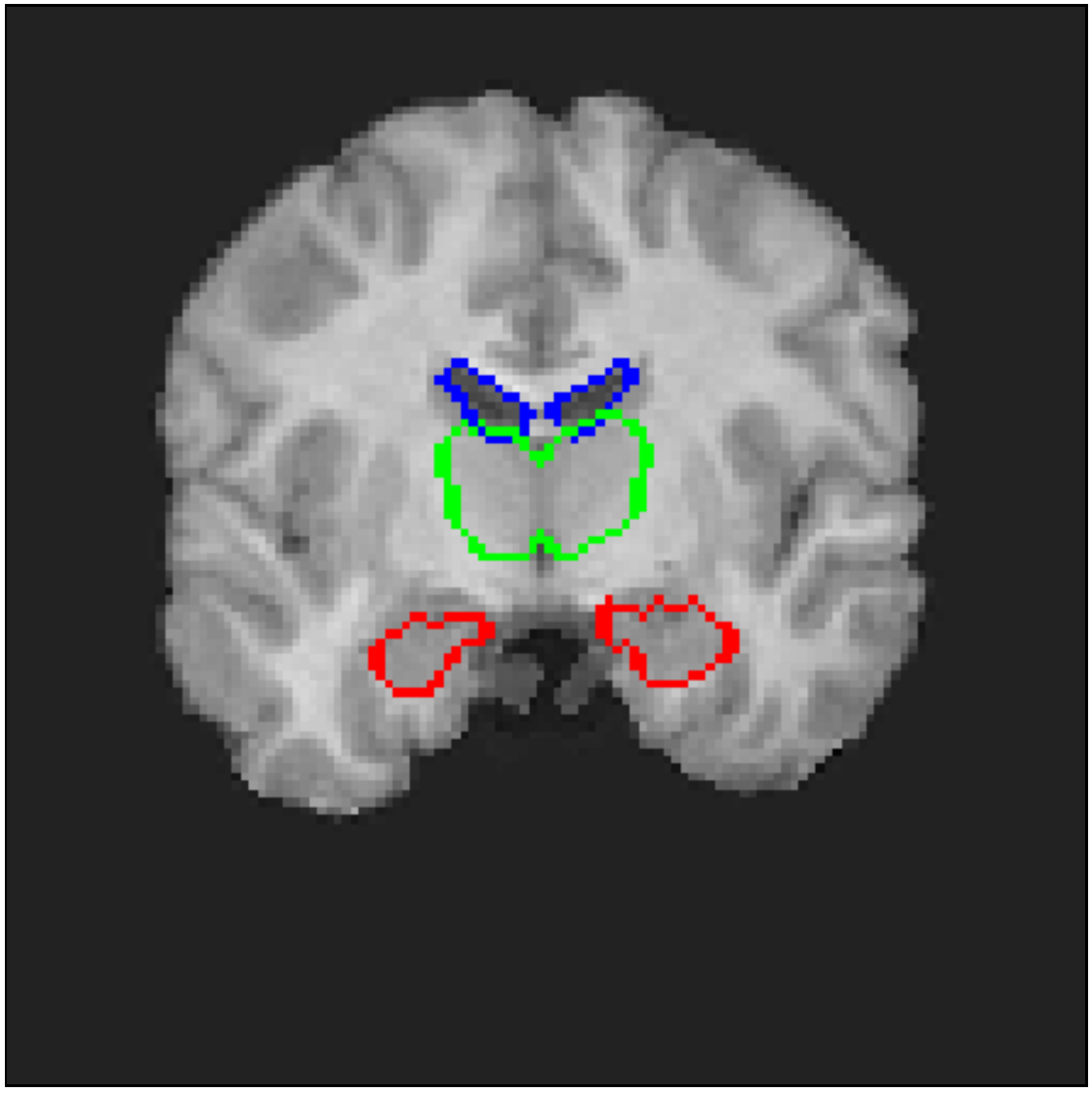}} &
                 \raisebox{-.5\height}{\includegraphics[height=0.16\textwidth]{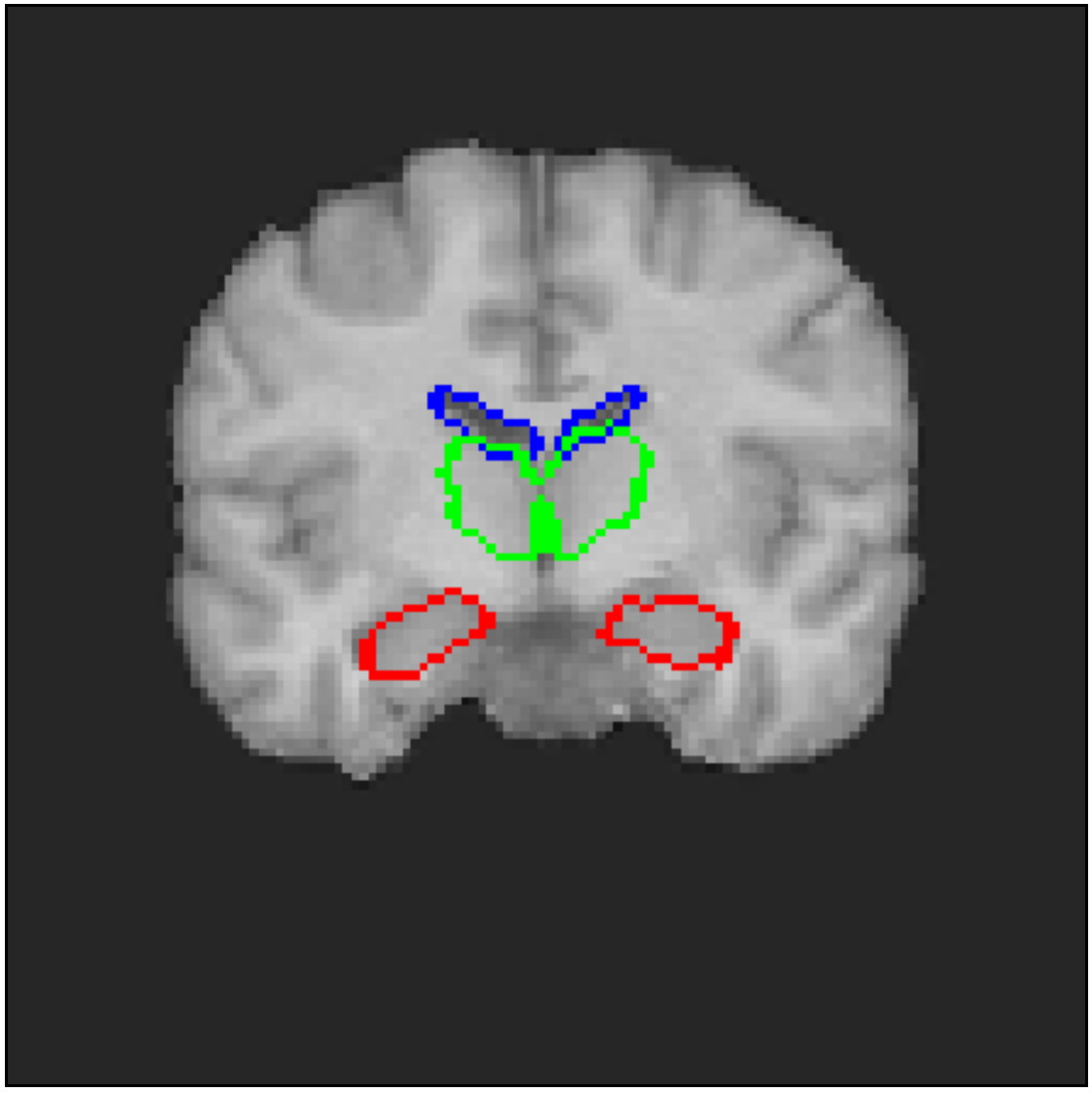}} &
                 \raisebox{-.5\height}{\includegraphics[height=0.16\textwidth]{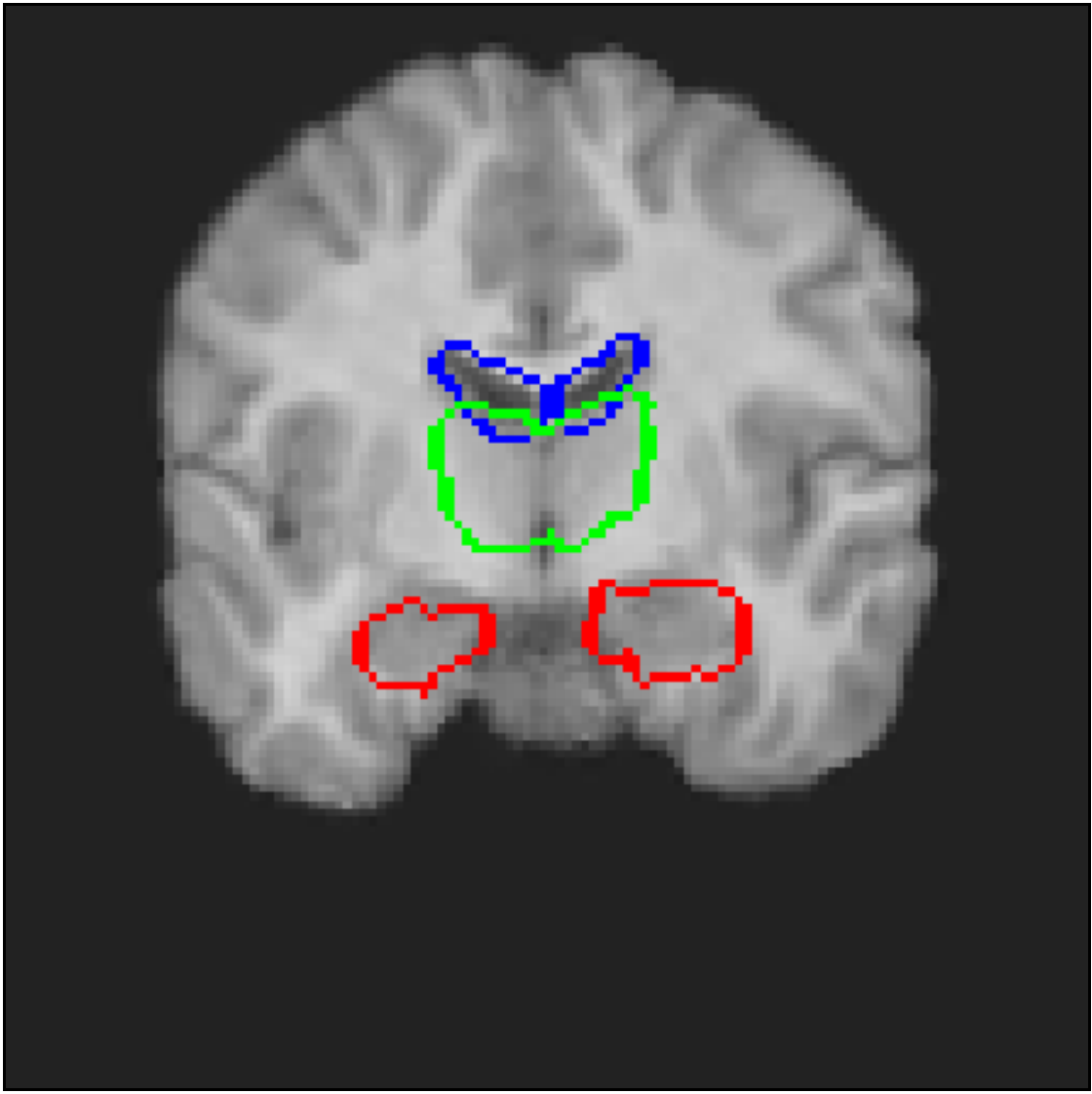}} &
                 \raisebox{-.5\height}{\includegraphics[height=0.16\textwidth]{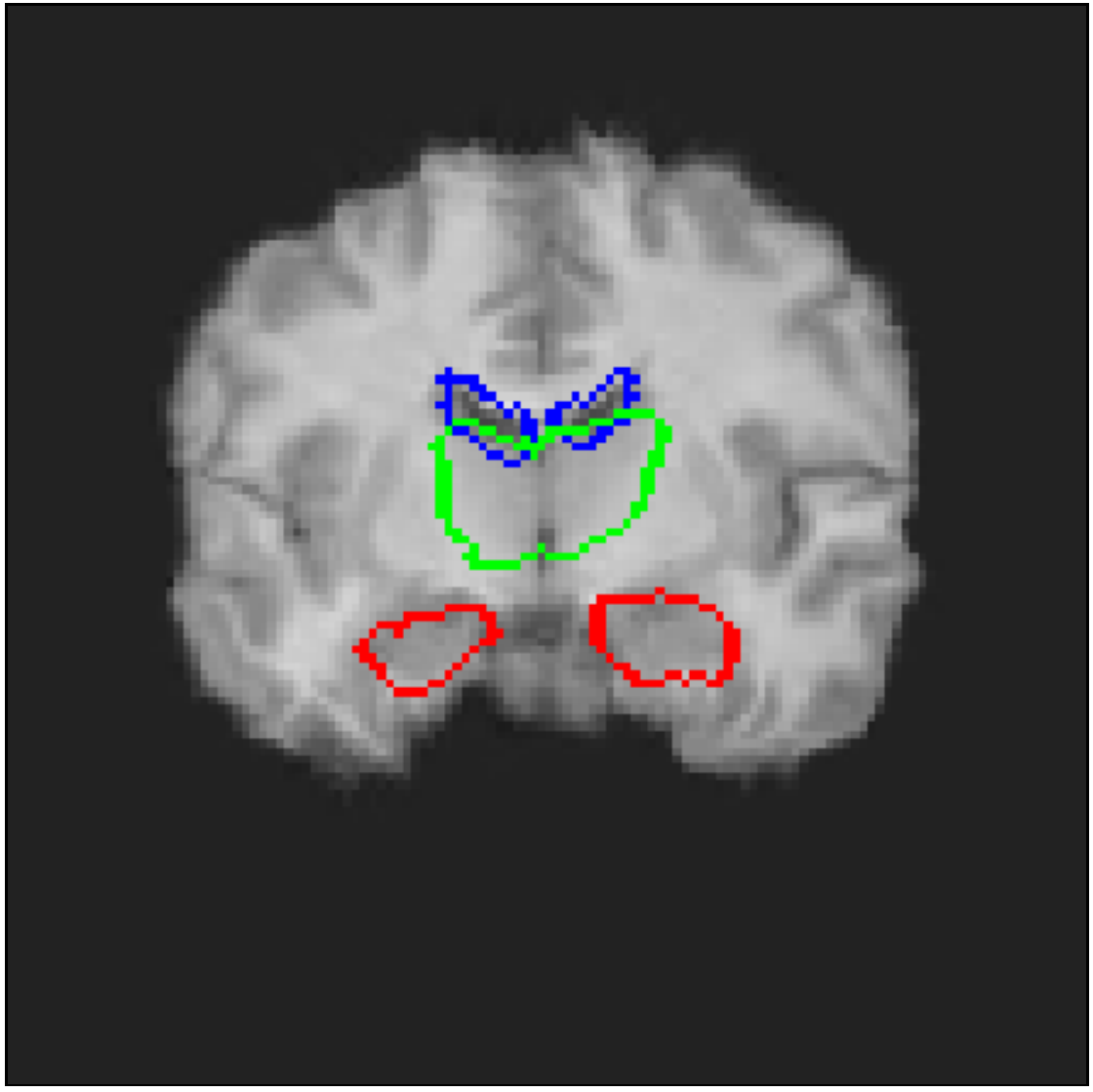}} &
                 \raisebox{-.5\height}{\includegraphics[height=0.16\textwidth]{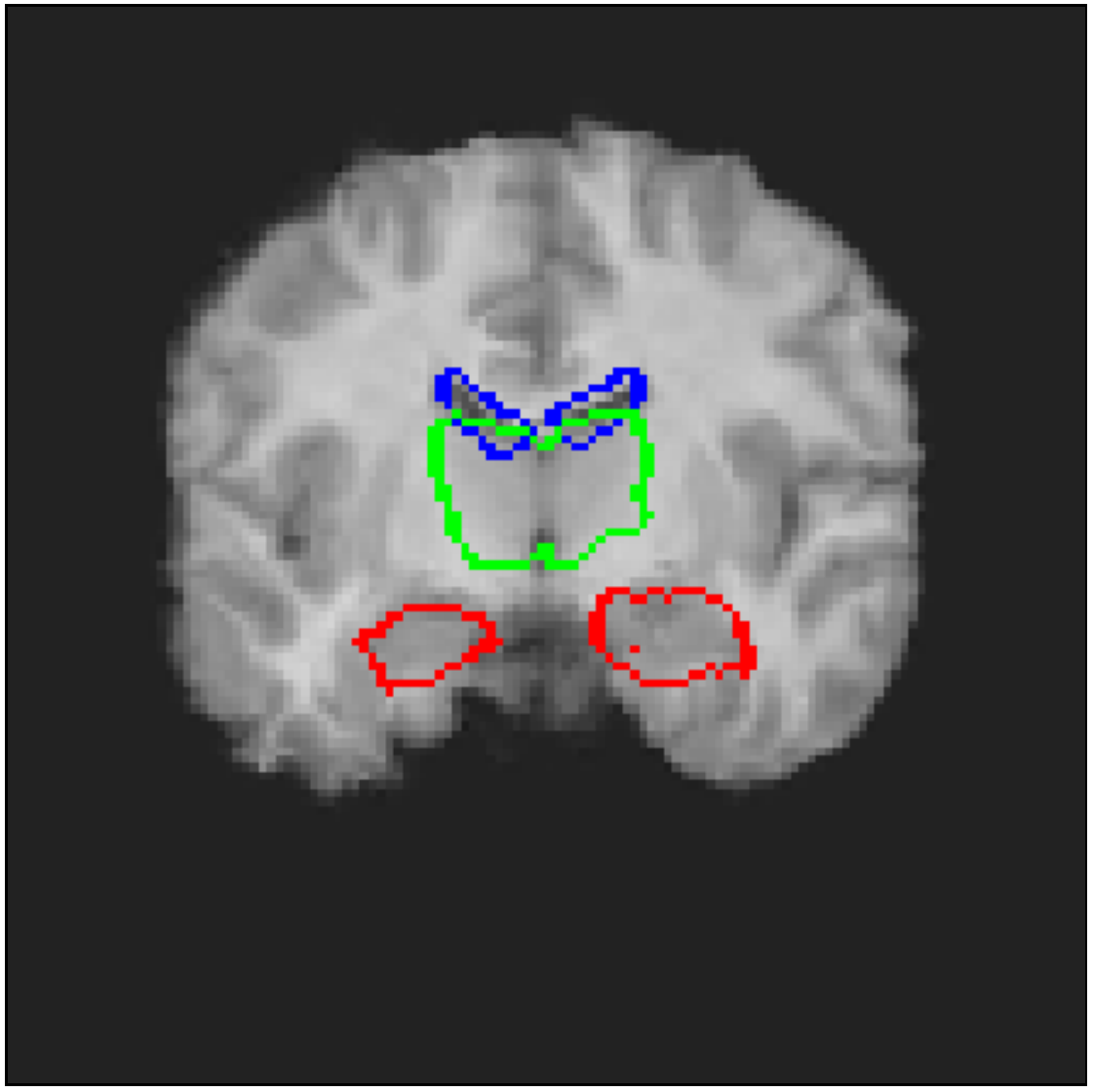}}
    \end{tabular}
   \caption{ IBSR18 sampled source, target, and warped images for SyN, VoxelMorph (VM), and our DDR-Net. Blue: ventricles, green: thalami, red: hippocampi.}
    \label{fig:comparison_seg_ibsr}
\end{figure}

\begin{table}[t]
\centering
\small
\begin{tabular}{l|l|c|cc}
\hline
\multirow{3}{*}{Dataset}&\multirow{3}{*}{Method}&\multirow{3}{*}{RMSE ($e^{-3}$)}&\multicolumn{2}{c}{Foldings}\\
\cline{4-5}&&&$|\sum det(J_{\phi})<0|$ & \, $\frac{1}{N}\sum \delta(det(J_{\phi})<0)$ \\
&&&$(\mathbf{e^{-5}})$ & \, (Ratio: $\mathbb{\permil}$)\\
\hline

\multirow{3}{*}{OASIS} 
& SyN~\cite{avants2011reproducible}  &  4.79 $\pm$ 0.001  & 495.37 $\pm$ 0.03 & 0.1 $\pm$ 0.768 \\ 
& VM~\cite{dalca2018unsupervised} & 1.25 $\pm$ 0.001 &  1.46 $\pm$ 0.005 & 0.051 $\pm$ 0.014 \\
& DDR-Net & \textbf{1.10 $\pm$ 0.001} & \textbf{1.09 $\pm$ 0.000}  &  \textbf{0.003 $\pm$ 0.019} \\ 
\hline

\multirow{3}{*}{IBSR18} 
& SyN~\cite{avants2011reproducible} & 10.93 $\pm$ 0.26 & \textbf{0.00 $\pm$ 0.000}  & \textbf{0.00 $\pm$ 0.00} \\ 
& VM~\cite{dalca2018unsupervised}  & 2.06 $\pm$ 0.001 &  20.17 $\pm$ 0.005 & 0.05 $\pm$ 0.014 \\ 
& DDR-Net  &\textbf{1.67 $\pm$ 0.001} & 17.266 $\pm$ 0.000 &  0.04 $\pm$ 0.117 \\ 
\hline

\multirow{3}{*}{3DIRCADB-01} 
& SyN~\cite{avants2011reproducible}  & 68.04 $\pm$ 0.024 & \textbf{476.98 $\pm$ 0.001} & 35.32 $\pm$ 5.227 \\ 
& VM~\cite{dalca2018unsupervised}  & 16.84 $\pm$ 0.005 & 1221.21 $\pm$ 0.01 & 0.72 $\pm$ 0.274 \\  
& DDR-Net  &\textbf{14.11 $\pm$ 0.004} & 723.89 $\pm$ 0.003 & \textbf{0.65 $\pm$ 0.247} \\ %
\hline
\end{tabular}
\caption{Comparison of the registration performance on all the three datasets.}
\vspace{-0.2in}
\label{tab:comparison_mse} 
\end{table}

\begin{table}[t]
\centering
\begin{tabular}{l|l|c|c|c}
\hline
{Dataset} & {Region} & {SyN~\cite{avants2008symmetric}} & {VM~\cite{dalca2018unsupervised}} & {DDR-Net} \\
\hline

\multirow{18}{*}{IBSR18} & 3rd ventricle & 0.28 $\pm$ 0.14 & 0.42 $\pm$ 0.16 & \textbf{0.44 $\pm$ 0.17} \\
& 4th ventricle & 0.20 $\pm$ 0.13 &	\textbf{0.39 $\pm$ 0.16} &	0.36 $\pm$ 0.16 \\
& amygdala &  \textbf{0.34 $\pm$ 0.13} & 0.27 $\pm$  0.23 & 0.31 $\pm$ 0.20 \\
& brainstem & 0.62 $\pm$ 0.12 &	\textbf{0.69 $\pm$  0.13} &	0.66 $\pm$  0.16  \\
& caudate &  \textbf{0.48 $\pm$ 0.09} & 0.40 $\pm$ 0.21 &	0.40 $\pm$ 0.20 \\
& cerebellum cortex & 0.55 $\pm$ 0.12 & 0.59 $\pm$ 0.20 & \textbf{0.63 $\pm$ 0.13} \\
& cerebellum white matter & 0.42 $\pm$ 0.14 &	0.34 $\pm$ 0.20 & \textbf{0.44 $\pm$ 0.18} \\
&cerebral cortex & 0.43 $\pm$ 0.09 & 0.55 $\pm$ 0.18 & \textbf{0.58 $\pm$ 0.13} \\
& cerebral white matter & 0.53 $\pm$ 0.06 & 0.58 $\pm$ 0.15 & \textbf{0.59 $\pm$ 0.11} \\
& csf & 0.22 $\pm$ 0.17 & 0.31 $\pm$ 0.16 & 0.31 $\pm$ 0.17 \\
& hippocampus & 0.32 $\pm$ 0.11 & 0.17 $\pm$ 0.14 & \textbf{0.34 $\pm$ 0.21}  \\
& lateral ventricle & 0.38 $\pm$ 0.12 & \textbf{0.53 $\pm$ 0.18} & 0.45 $\pm$ 0.17  \\
& pallidum & 0.24 $\pm$ 0.16 & 0.23 $\pm$ 0.23 & \textbf{0.29 $\pm$ 0.24} \\
& putamen	& 0.29 $\pm$ 0.21 & 0.27 $\pm$ 0.25 & \textbf{0.33 $\pm$ 0.27} \\
& thalamus & \textbf{0.68 $\pm$ 0.08} & 0.61 $\pm$ 0.18 & 0.60 $\pm$ 0.19 \\
& ventraldc &  0.49 $\pm$ 0.11 &	0.49 $\pm$ 0.21 & \textbf{0.54 $\pm$ 0.18}  \\

&\textbf{Avg. Dice} & 0.28 $\pm$ 0.14 &	0.43 $\pm$ 0.1 & \textbf{0.47 $\pm$ 0.1}  \\
\hline
\multirow{3}{*}{3DIRCADB-01} & bone & 0.223 $\pm$ 0.106 & 0.356 $\pm$ 0.149 & \textbf{0.369 $\pm$ 0.153} \\
& skin & 0.889 $\pm$ 0.065 &	0.982 $\pm$ 0.012 &	\textbf{0.991 $\pm$ 0.009} \\
& liver & 0.690 $\pm$ 0.099 & 0.731 $\pm$  0.122 & \textbf{0.741 $\pm$ 0.131} \\
\hline
\end{tabular}
\caption{Segmentation comparison  using SyN, VoxelMorph (VM) and DDR-Net. }\label{tab:dice_comp} 
\vspace{-0.2in}
\end{table}

\section{Experiments}
\vspace{0.1in}
\noindent
We evaluate our method on three public datasets, which involves two types of 3D medical images, i.e., brain MRI scans and liver CT scans. 

\vspace{0.05in}
\noindent
\textbf{OASIS Dataset~\cite{marcus2007open}.} The T1-weighted brain scans from the OASIS dataset underwent pre-processing steps including down-sampling, skull-stripping, intensity normalization to the range [0, 1], bias field correction, and co-registration with affine transformations. We have resulting images of resampled resolution $128 \times 128 \times 128$ and the voxel size of $1.25 \times 1.25 \times 1.25 mm^3$. We collect 360 volume pairs for the 3D experiments.

\vspace{0.05in}
\noindent
\textbf{IBSR18 Dataset~\cite{valverde2015comparison}.} The IBSR18 dataset consists of T1-weighted scans for 18 subjects with dimensions of $256 \times 128 \times 256$. We resampled these images to $128 \times 128 \times 128$. Scans underwent preprocessing steps including skull-stripping, bias field correction, and intensity normalization. Each scan also comes with 84 manually labelled anatomical structures. We collect 306 3D image pairs. Segmentation maps including 28 anatomical structures (see Table~\ref{tab:dice_comp}) were also obtained in this dataset, which followed the same preprocessing steps of resampling. 

\vspace{0.05in}
\noindent
\textbf{3DIRCADB-01 Dataset~\cite{3dircadb}.} The 3DIRCADB-01 dataset contains 20 CT scans with masks of the segmented structures available for bone, liver, and skin. Scans underwent preprocessing steps including histogram equalization and intensity normalization to the range [0, 1]. We collect and pair the central slices on the axial plan and obtained 380 2D image pairs with a resolution of $512 \times 512$.

\vspace{0.05in}
\noindent
\textbf{Experimental Settings.} For all the experiments, we use 70\% of the collected data for training, 10\% for validation, and 20\% for test, after a random shuffle. We use the Adam \cite{kingma2014adam} optimizer with a learning rate of $e^{-4}$. All the experiments were trained using NVIDIA GeForce TITAN X GPUs. 

We measure the image matching after registration using the intensity root mean square error (RMSE), and the smoothness of a deformation by counting the number of its foldings and the absolute sum of negative determinant of its Jacobian~\cite{ashburner2007fast}. Also, we perform the segmentation task using image registration and report the Dice score in terms of volume overlap. 



We compare our approach with ANTsPy package using Symmetric Normalization (SyN)~\cite{avants2011reproducible}. We use cross-correlation and other default settings, which are optimal for our task with 201 iterations for registration. Another baseline algorithm is VoxelMorph~\cite{dalca2018unsupervised}, and we also use its default settings for comparison. Initial affine registration is not performed for any of the experiments. 

\vspace{0.05in}
\noindent
\textbf{Experimental Results.} Figures~\ref{fig:comparison_oasis3d},~\ref{fig:comparison_ibsr3d},~\ref{fig:comparison_3dircadb} show qualitative results for OASIS, IBSR18, and 3DIRCADB-01 datasets, respectively. Overall, our approach produces better matching results as compared to the baseline methods, demonstrated by the image difference plots. VoxelMorph tends to produce unwanted artifacts in its deformed images, which are not seen in our results. The deformation plot and the visualization of the determinant of the Jacobian show that our results have less deformations happening in the background, as expected. Also, our method achieves significantly smaller deformations throughout the image space and provides a better matching result to the target at the same time. 

We observe that the VoxelMorph produces more irregular deformation fields compared to the other methods. Table~\ref{tab:comparison_mse} show the quantitative analysis of the registration performance on the OASIS, IBSR18, and 3DIRCADB-01 datasets. We achieve consistently lower number of foldings compared to the baseline methods while maintaining the consistently higher registration accuracy.
Table~\ref{tab:dice_comp} indicates the Dice score of anatomical structures for the IBSR18 and 3DIRACADB-01 datasets. Our method performs better on most of the anatomical structures for both the IBSR18 and 3DIRCADB-01 datasets.  Figure~\ref{fig:comparison_seg_ibsr} shows the segmentation maps for a few anatomical structures from the IBSR18 dataset. 

Regarding the inference time, ANTs SyN, which is tested on CPU since it does not have a GPU implementation, takes on an average 20 minutes, whereas VoxelMorph tested on GPU takes 121 milliseconds, while DDR-Net tested on GPU takes 425 milliseconds to register an image pair from the OASIS dataset. 
A limited ablation study conducted by us revealed that a single scale UNet architecture working on the full size velocity field integration for 2D input size of $128 \times 128$ for a single input image pair takes 26.46 MB of memory, whereas an architecture like DDR-Net for the same input and UNet architecture will take 21.16 MB of memory. The memory requirement was calculated considering the forward pass and parameters in the neural network. For 3D cases, our DDR-Net can take in more images in a batch, compared to a single scale UNet.



\section{Conclusion and Discussion}
In this paper, we have proposed a diffeomorphic image registration model which estimates smoother deformations and provides a better image matching result. It leverages both the global context and local fine structures of the data effectively to enhance the registration result and handle large deformations as a result of such an architecture. Generally, our approach produced more regular deformation fields, which are significantly smoother than the baseline methods. The dice scores indicate that our approach is comparable to the existing methods even surpassing them in most cases. Moreover, our architecture shows that simple yet efficient changes in architectures can lead to better deep learning strategies.
We believe that an optimal balance of GPU memory and accuracy is essential for registering image pairs in 3D in order to maximize the utilization of 3D information. Our work will push new avenues of research in considering memory efficient registration models in deep learning, which fully leverage the data.

 \bibliographystyle{splncs04}
 \bibliography{meta.bib}

\begin{thebibliography}{10}
\providecommand{\url}[1]{\texttt{#1}}
\providecommand{\urlprefix}{URL }
\providecommand{\doi}[1]{https://doi.org/#1}

\bibitem{3dircadb}
3dircadb-01. \url{https://www.ircad.fr/research/3dircadb/}

\bibitem{arsigny2006log}
Arsigny, V., Commowick, O., Pennec, X., Ayache, N.: A log-euclidean framework
  for statistics on diffeomorphisms. In: International Conference on Medical
  Image Computing and Computer-Assisted Intervention. pp. 924--931. Springer
  (2006)

\bibitem{ashburner2007fast}
Ashburner, J.: A fast diffeomorphic image registration algorithm. Neuroimage
  \textbf{38}(1),  95--113 (2007)

\bibitem{avants2008symmetric}
Avants, B.B., Epstein, C.L., Grossman, M., Gee, J.C.: Symmetric diffeomorphic
  image registration with cross-correlation: evaluating automated labeling of
  elderly and neurodegenerative brain. Medical image analysis  \textbf{12}(1),
  26--41 (2008)

\bibitem{avants2011reproducible}
Avants, B.B., Tustison, N.J., Song, G., Cook, P.A., Klein, A., Gee, J.C.: A
  reproducible evaluation of ants similarity metric performance in brain image
  registration. Neuroimage  \textbf{54}(3),  2033--2044 (2011)

\bibitem{beg2005computing}
Beg, M.F., Miller, M.I., Trouv{\'e}, A., Younes, L.: Computing large
  deformation metric mappings via geodesic flows of diffeomorphisms.
  International journal of computer vision  \textbf{61}(2),  139--157 (2005)

\bibitem{dalca2018unsupervised}
Dalca, A.V., Balakrishnan, G., Guttag, J., Sabuncu, M.R.: Unsupervised learning
  for fast probabilistic diffeomorphic registration. In: International
  Conference on Medical Image Computing and Computer-Assisted Intervention. pp.
  729--738. Springer (2018)

\bibitem{hering2019mlvirnet}
Hering, A., van Ginneken, B., Heldmann, S.: mlvirnet: Multilevel variational
  image registration network. In: International Conference on Medical Image
  Computing and Computer-Assisted Intervention. pp. 257--265. Springer (2019)

\bibitem{higham2005scaling}
Higham, N.J.: The scaling and squaring method for the matrix exponential
  revisited. SIAM Journal on Matrix Analysis and Applications  \textbf{26}(4),
  1179--1193 (2005)

\bibitem{kingma2014adam}
Kingma, D.P., Ba, J.: Adam: A method for stochastic optimization. arXiv
  preprint arXiv:1412.6980  (2014)

\bibitem{krebs2019learning}
Krebs, J., Delingette, H., Mailh{\'e}, B., Ayache, N., Mansi, T.: Learning a
  probabilistic model for diffeomorphic registration. IEEE transactions on
  medical imaging  \textbf{38}(9),  2165--2176 (2019)

\bibitem{krebs2018unsupervised}
Krebs, J., Mansi, T., Mailh{\'e}, B., Ayache, N., Delingette, H.: Unsupervised
  probabilistic deformation modeling for robust diffeomorphic registration. In:
  Deep Learning in Medical Image Analysis and Multimodal Learning for Clinical
  Decision Support, pp. 101--109. Springer (2018)

\bibitem{marcus2007open}
Marcus, D.S., Wang, T.H., Parker, J., Csernansky, J.G., Morris, J.C., Buckner,
  R.L.: Open access series of imaging studies (oasis): cross-sectional mri data
  in young, middle aged, nondemented, and demented older adults. Journal of
  cognitive neuroscience  \textbf{19}(9),  1498--1507 (2007)

\bibitem{ronneberger2015u}
Ronneberger, O., Fischer, P., Brox, T.: U-net: Convolutional networks for
  biomedical image segmentation. In: International Conference on Medical image
  computing and computer-assisted intervention. pp. 234--241. Springer (2015)

\bibitem{valverde2015comparison}
Valverde, S., Oliver, A., Cabezas, M., Roura, E., Llad{\'o}, X.: Comparison of
  10 brain tissue segmentation methods using revisited ibsr annotations.
  Journal of Magnetic Resonance Imaging  \textbf{41}(1),  93--101 (2015)

\bibitem{yang2017quicksilver}
Yang, X., Kwitt, R., Styner, M., Niethammer, M.: Quicksilver: Fast predictive
  image registration--a deep learning approach. NeuroImage  \textbf{158},
  378--396 (2017)

\bibitem{zhang2015finite}
Zhang, M., Fletcher, P.T.: Finite-dimensional lie algebras for fast
  diffeomorphic image registration. In: International conference on information
  processing in medical imaging. pp. 249--260. Springer (2015)

\end{thebibliography}

\end{document}